\documentclass[a4paper,11pt]{article}

\usepackage{jcappub}
\usepackage{latexsym,amsmath,amssymb,hyperref}
\usepackage{slashed}
\usepackage{tensor}

\usepackage[T1]{fontenc} 

\usepackage[utf8]{inputenc}

\usepackage[T1]{fontenc} 

\usepackage{Extrapackages}
\usepackage{xcolor}
\usepackage{cancel}
\usepackage{tensor}
 \usepackage{graphicx}
\usepackage{simpler-wick}
\usepackage{tikz-feynman}
\tikzfeynmanset{compat=1.0.0}
\newcommand{\cR}{\mathcal{R}}
\newcommand{\cK}{\mathcal{K}}
\newcommand{\cT}{\mathcal{T}}
\newcommand{\cO}{\mathcal{O}}
\newcommand{\q}{\textbf{k}}

\begin{document}

\title{ \begin{center}
		\boldmath{Non-Gaussianities in models of inflation with large and negative entropic masses}
\end{center} }

\author{Ricardo Z.~Ferreira}

\affiliation{Nordita, KTH Royal Institute of Technology and Stockholm University, \\ Roslagstullsbacken 23, SE-106 91 Stockholm, Sweden}

\emailAdd{ricardo.zambujal@su.se}

\abstract{Models of inflation where the entropic directions have large and negative masses $|m_s| \gg H$ can be well described by a single-field EFT with an imaginary sound speed $c_s$.  Among other features, they predict an exponential enhancement of the spectrum of scalar perturbations which however is not inherited by non-Gaussianities. 
	In this work, I complete the calculation of the trispectrum in this EFT by considering the contributions from the contact interaction and the exchange diagram. While for most shapes the trispectrum is approximately constant, I find that for certain configurations where all the momenta collapse to a line the trispectrum is proportional to $(m_s/H)^5$ for the contact interaction and to $(m_s/H)^6$ for the exchange diagram, as anticipated by previous work. I also discuss the UV sensitivity of the results and argue why the EFT provides a good order of magnitude estimate. 
	In the end, I confront the different predictions for the scalar spectrum against observations. In models where the entropic mass is proportional to a positive power of the slow-roll parameter $\epsilon$, like in hyperinflation, the spectrum grows on small scales and becomes constrained by the overproduction of primordial black holes. Imposing such constraint jointly with the correct amplitude and spectral tilt at CMB scales excludes a large set of potentials.  Only those where the spectral tilt is controlled by $  m_s \delta/H \sim {\cal O}(-0.01)$, where $\delta=\dot{\epsilon}/(\epsilon H)$ is the second slow-roll parameter, are likely observationally viable. 
	Finally, the constraints on the bispectrum generically impose $|c_s m_s|/H \lesssim 10-20$ while those on the trispectrum give a weaker bound when using the constraints on $g_\text{NL}^{ \dot{\sigma}^4}$ as a proxy. For hyperinflation the bispectrum bound translates into $\omega \lesssim 11$ where $\omega$ is the turning rate in field-space.
}
\maketitle
\flushbottom

\section{Introduction}

In the inflationary paradigm, what protects the inflaton from large radiative corrections and a new naturalness problem?
Symmetries are naturally good mass protectors and have been thoroughly explored, for example, in the case of shift symmetries \cite{Freese:1990rb,Silverstein:2008sg,Kaloper:2008fb}. However, the latest observational data does not seem to point in that direction, at least for the minimal scenarios \cite{Akrami:2018odb}. Moreover, the symmetry breaking scale generating such Goldstone boson tends to be (super) Planckian thus raising concerns about large gravitational corrections (see \cite{Hebecker:2019vyf} for a recent discussion).

Another interesting possibility is to have a dynamical mechanism where multi-field effects counteract the rolling of the inflaton thus effectively allowing for inflation in steeper potentials. In that regard, considerable attention has been given to multi-fields models with a strongly non-geodesic motion (see \cite{Cremonini:2010ua,Achucarro:2010jv,Achucarro:2012yr,Cespedes:2012hu,Renaux-Petel:2015mga} for a few examples). In particular, more recently it has been noted that even when the field space has a negative curvature $R_\text{fs}$, of which hyperbolic geometries are prototypical examples \cite{Renaux-Petel:2015mga,Brown:2017osf,Mizuno:2017idt,Christodoulidis:2018qdw,Garcia-Saenz:2018ifx,Bravo:2019xdo,Aragam:2019khr,Garcia-Saenz:2019njm,Bjorkmo:2019aev,Bjorkmo:2019fls, Chakraborty:2019dfh,Christodoulidis:2019jsx}, new slow-roll attractors in the inflationary trajectory are still possible to find. In these attractors, which have been argued to belong all to the same class \cite{Bjorkmo:2019fls}, the initially spectator field is typically displaced from its minima and there is a fast turning rate in field-space.

Parallelly to this discussion, it is known that for any two-field system the quadratic action for perturbations assumes a very simple form where the shift symmetric adiabatic curvature perturbation, $\zeta$, couples derivatively to the entropic degree of freedom whose effective mass  is controlled by  \cite{Langlois:2008mn}
\begin{eqnarray} \label{entropic mass}
m_s^2 =V_{;ss} + H^2 ( \epsilon R_\text{fs} M_p^2 - \eta_\perp^2 )  \, ,
\end{eqnarray}
where $V$ is the multi-field potential and $\eta_\perp = - V_{,s}/(H \dot{\phi})$ controls the bending of the trajectory in field-space. The covariant derivatives are taken with respect to the field-space metric and the subscript $s$ denotes the entropic direction. 
As usual, when the entropic direction is massive it can be integrated out yielding a single-field effective field theory (EFT) with a reduced speed of sound \cite{Cheung:2007st,Achucarro:2010da,Achucarro:2012sm}. However, recently it has been pointed out that even if $m_s^2$ is large and negative, $-m_s^2/H^2 \gg 1$, the entropic direction can still be integrated out using its equation of motion \cite{Garcia-Saenz:2018ifx}. The resulting theory, valid for wavelengths smaller than $|m_s|$, is still a single-field EFT but with an imaginary sound speed controlled by $c_s^{-2}= 1+4 \eta_\perp^2 H^2/ m_s^2<0$.
Remarkably, even if $\eta_\perp^2 \gg 1$ dominates the effective mass $m_s$ making it large and negative, the theory is still stable because on superhorizon scales the effective entropic mass is instead $m_s^2+4 \eta_\perp^2 H^2>0$. 
The consequences though are rather non-trivial. The temporary negative mass causes an exponential growth of $\zeta$ for modes in the window $H \lesssim k/a \lesssim |m_s|$, similarly to what happens in the context of axial couplings to gauge fields during inflation \cite{Anber2010,Barnaby2011,Ferreira2014,Ferreira2017}.

Non-Gaussianities are another feature of the models described by such EFT. Interestingly, they do not inherit the exponential growth but instead a polynomial dependence on $m_s/H$. This has been shown for the 3-point function \cite{Garcia-Saenz:2018vqf} and also for higher-order correlators \cite{Bjorkmo:2019qno,Fumagalli:2019noh}. Namely, while for most shapes the non-Gaussian parameters are constant in the $m_s \gg H$ limit, for certain shapes approaching the collapsing limit, where all momenta in the correlator collapse to a line (see fig. \ref{fig:flat3pf}), the time integrals become dominated by the UV cutoff with a leading power-law dependence of $\alpha^3$ in the bispectrum, and $\alpha^6$ in the trispectrum where $\alpha \equiv - |c_s m_s|^2/H^2$ \cite{Garcia-Saenz:2018vqf,Fumagalli:2019noh}. The peak on flattened shapes and its UV sensitivity are typical of theories with excited initial states \cite{Chen:2006nt,Chen:2009bc}. It shows that to find a precise result for such shapes one necessarily needs to consider the full multi-field system. However, as I will discuss later on, if the theory quickly approaches that of approximately massless and weakly coupled scalar fields for scales above the cutoff, the UV contribution will be strongly suppressed thus validating the order of magnitude estimation within the EFT.

In this work, I will complete the trispectrum computation by evaluating the contributions from the contact interaction and the exchange diagram (see fig. \ref{fig:4ptX}). So far only one term of the exchange diagram has been computed \cite{Fumagalli:2019noh}. I find results consistent with what has been found in the literature, i.e., that the exchange diagram is the dominant contribution to the trispectrum and that it peaks on shapes where all the momenta are collinear. Afterwards, I will collect the different predictions of these models for the spectrum of scalar perturbations: amplitude, spectral tilt and non-Gaussianity; and confront them against observations. In some cases, I will make quantitative assessments on the observational viability of the theory while in other the model dependence only allows for qualitative statements.

The paper is organized as follows. In sec. \ref{sec:Basics of the model} I give the basic ingredients of the EFT with imaginary sound speed. In sec. \ref{sec:non-Gaussianities} I look at the non-Gaussianities, first, by revisiting the bispectrum in sec. \ref{sec:bispectrum} and then by evaluating the trispectrum in sec. \ref{sec:trispectrum}. In sec. \ref{observational constraints} I use observations to place constraints on the models. Auxiliary and intermediate expressions are presented in the appendix. Namely, app. \ref{app: rapid-turn attractors} briefly discusses 2 rapid-turn attractor models, app. \ref{app:interaction Hamiltonians } presents the interaction Hamiltonians and sec. \ref{app:equilateral trispectrum} a few expressions related to the trispectrum including the results for the equilateral shape.

\section{Basics of the model \label{sec:Basics of the model}}

The models under discussion are those where the entropic direction\footnote{These statements can be generalized to a system with several entropic directions as long as they are all large and negative \cite{Bjorkmo:2019aev}.} has an effective mass $m_s^2$ which is large and negative, $ -m_s^2/H^2 \equiv  \alpha^2/|c_s|^2  \gg 1$. In such cases, one can use the single-field EFT with imaginary sound speed to compute non-Gaussianities  \cite{Garcia-Saenz:2018vqf}. In this section, I review the basic features of the EFT and set the notation for the following sections.

 To see why the EFT is a good description let me consider a model with 2 fields  $\phi_1,\phi_2$ described by the Lagrangian
\begin{eqnarray}
{\cal L} = - \frac 1 2 G_{ab} \partial_\mu \phi^a  \partial^\mu \phi^b  - V \, ,\label{background lagrangian}
\end{eqnarray}
where $G_{ab}(\phi_1,\phi_2) $ is the field-space metric and $V(\phi_1,\phi_2) $ the potential energy. Both $G_{ab}$ and $V$ are chosen such that the background is in slow-roll, either the standard slow-roll attractor or a different one.
Remarkably, at the level of perturbations, the Lagrangian at quadratic order is generically given by \cite{Langlois:2008mn}
\begin{eqnarray}
{\cal L} = - \epsilon M_p^2 (\partial \zeta)^2 - \frac 1 2 (\partial \sigma)^2 - \frac 1 2 m_s^2 \sigma^2 - 2 \dot{\phi} \eta_\perp \sigma \dot{\zeta}^2  \, , \label{Lagrangian for the quadratic perturbations}
\end{eqnarray}
where $\zeta$ is the adiabatic curvature perturbation, $\dot{\phi}=(G_{ab}\dot{\phi}^a\dot{\phi}^b)^{1/2}$ is the background velocity in the adiabatic direction and $\sigma$ the entropic field, i.e. the perturbation orthogonal to the background trajectory. The Lagrangian in eq. \eqref{Lagrangian for the quadratic perturbations} is characterized by 3 parameters: the slow-roll parameter  $\epsilon=-\dot{H}^2/H^2$, the effective entropic mass $m_s$ defined in eq. \eqref{entropic mass} and $\eta_\perp$.
As usual, when $|m_s^2| \gg H^2$ one can integrate out $\sigma $ and define an EFT for modes below the cutoff $k/a \lesssim |m_s|$. Namely, for those modes, the equation of motion for $\sigma$ reads
\begin{eqnarray}
\sigma \simeq -2 \frac{\dot{\phi}\eta_\perp}{m_s^2} \dot{\zeta} \, .
\end{eqnarray} 
Substituting this equality in eq. \eqref{Lagrangian for the quadratic perturbations} then yields the single-field EFT \cite{Christodoulidis:2018qdw,Garcia-Saenz:2018vqf}
\begin{eqnarray}
{\cal L} = \epsilon M_p^2 \left[ \frac{\dot{\zeta}^2}{c_s^2}  +k^2 \zeta^2\right] \, ,
	\end{eqnarray}
where $c_s^{-2} = 1+4 \eta_\perp^2 H^2/ m_s^2$ is the effective speed of sound. If $m_s^2+4 \eta_\perp^2 H^2>0$  the entropic direction decays on superhorizon scales thus ensuring the stability of the system. However, the effective speed of sound becomes imaginary, $c_s^2<0$, leading to an exponential growth in the time interval $H \lesssim |c_s| k/a \lesssim |m_s|$. This can be seen more explicitly by solving the equation of motion for $\zeta$ in the constant $c_s^2$ approximation which yields\footnote{Note that I added an extra (-) sign compared to the solution in \cite{Garcia-Saenz:2018vqf} in order to make the real and imaginary parts of $\zeta$ positive on superhorizon scales.} \cite{Garcia-Saenz:2018vqf}
\begin{equation} \label{mode function}
\zeta_{k}(\tau)=\frac{\beta e^{2x}}{k^{3/2}} \left[-f(k,\tau)+\rho e^{i \psi-2 x } g(k, \tau) \right]
\end{equation}
where $f$ and $g$ are real functions given by
\begin{eqnarray} \label{g and f}
f(k,\tau)= e^{k\left|c_{s}\right| \tau}\left(k\left|c_{s}\right| \tau-1\right) \, , \qquad 
g(k, \tau)= - f(k,-\tau) \, ,
\end{eqnarray}
and $x, \beta, \rho, \psi$ are real parameters that need to be computed in the full theory which I assume to be approximately scale-invariant.
On super horizon scales, for $\rho e^{-2x} \ll 1$ the power spectrum is basically determined by the first term in eq. \eqref{mode function}
\begin{eqnarray} \label{power spectrum}
P_\zeta \simeq \frac{\beta^2 e^{2x}}{2\pi^2} \, .
\end{eqnarray}
By imposing the Wronskian condition on the solution in eq. \eqref{mode function} one further finds the relation \cite{Garcia-Saenz:2018vqf} 
\begin{eqnarray} \label{beta squared}
\beta^2 \rho \sin(\theta)= \pi^2 P_{\zeta,0}
\end{eqnarray}
where $P_{\zeta,0}= H^2/(8\pi^2 \epsilon |c_s|M_p^2)$ is the power spectrum in standard single-field inflation. Fortunately, the leading contributions to non-Gaussianities in the large $x$ limit are proportional to powers of $\beta^2 \rho \sin(\theta)$. Therefore, the results will only depend only on $\alpha$ and $|c_s|$. In light of that, and without loss of generality, I fix $\rho=1, \psi=\pi/2$ and use instead
\begin{equation}
\zeta_{k}(\tau)=\sqrt{\frac{2 \pi^{2} P_\zeta}{k^{3}}} \left[-f(k,\tau)+i e^{-2 x } g(k, \tau) \right] \label{mode function 2}
\end{equation}
 with $P_\zeta = P_{\zeta,0} e^{2 x}/2$. The parameter $x$ cannot be computed within the EFT. However, if all fields become approximately massless and weakly coupled on scales smaller than the cutoff $|c_s| k/a \lesssim |m_s|$, one expects to recover plane wave solutions at those scales and so to have $x \sim \alpha$. Nevertheless, the precise relation between $x, \alpha$ and $c_s$ will depend on the model. For example, in the case of rapid-turn attractors (see app. \ref{app: rapid-turn attractors}), where the turning rate in field-space is $\omega$ and the entropic mass is $m_s^2\simeq (\xi -1) \omega^2 H^2$, numerical and analytical studies have shown that $x \simeq (2- \sqrt{3+\xi} ) \pi \omega/2$ and $|c_s|=\sqrt{(1-\xi)/(3+\xi)}$ for $\xi <1$ \cite{Mizuno:2017idt,Bjorkmo:2019qno}. In particular, for hyperinflation $c_s^2=\xi=-1$ \cite{Brown:2017osf}.

Before proceeding to the computation of non-Gaussianities it is useful to introduce the variable $\cR$ defined as
\begin{eqnarray} \label{cR definition}
\mathcal{R} \equiv  \frac{\zeta}{\sqrt{ \pi^2 P_{\zeta,0}}} \, , \qquad 	\left< \mathcal{R}_k \mathcal{R}_q \right> = \frac{2P_\zeta}{ P_{\zeta,0} k^3} \delta^{(3)}(k+q) \, , \label{Dimensionless variable} 
\end{eqnarray}
which allows to directly extract the relevant quantities from the mode functions and from the interaction Hamiltonian.
In particular note that on superhorizon scales 
\begin{eqnarray} \label{real and imaginary part of R}
\text{Re} \left[\cR_k(0)\right]= \sqrt{\frac{2P_\zeta}{P_{\zeta,0}k^3}} \, , \quad \quad
\text{Im} \left[\cR_k(0)\right]= \sqrt{\frac{P_{\zeta,0}}{2P_\zeta k^3}} \, .
\end{eqnarray}

\section{Non-Gaussianities \label{sec:non-Gaussianities}}

\begin{figure}
	\centering
	\includegraphics[width=0.48\linewidth]{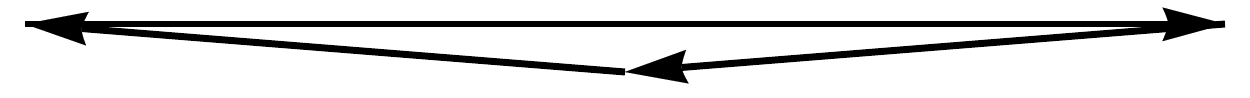}
	\includegraphics[width=0.48\linewidth]{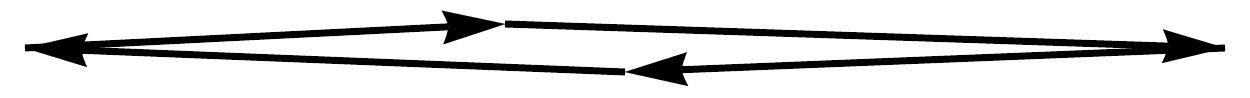}
	\caption{Flattened configurations}
	\label{fig:flat3pf}
\end{figure}

The shape and size of non-Gaussianities are intrinsic characteristics of a given inflationary model. Interestingly, in the context of the EFT described in the previous section it has been shown that the non-Gaussian parameters have a polynomial, rather than exponential, dependence on $m_s$ in the $|m_s| \gg H$ limit, or equivalently, when $x,\alpha \gg 1$ \footnote{I will always work in this limit throughout this section. The subleading corrections in $x$ are suppressed by $e^{-2x}$ while the subleading terms in $\alpha$ are suppressed by $\alpha^{-1}$.} \cite{Garcia-Saenz:2018vqf,Fumagalli:2019noh,Bjorkmo:2019qno}. Moreover, they peak in shapes where all the momenta collapse to a line (see fig. \ref{fig:flat3pf}). This characteristic, typical of theories with excited initial states \cite{Chen:2006nt,Chen:2009bc}, is easy to understand by looking at the time integrals in the in-in formalism. More concretely, the n-point function is given by \cite{Calzetta1987,Weinberg2005} 
\begin{eqnarray} \label{in-in expansion}
\left\langle\zeta^{n}(\tau)\right\rangle=\sum_{k=0}^{\infty} i^{k} \int_{-\infty}^{\tau} d \tau_{1} \ldots \int_{-\infty}^{\tau_{k-1}} d \tau_{k}\left\langle\left[H_{I}\left(\tau_{k}\right), \ldots\left[H_I \left(\tau_{1}\right), \zeta^{n}(\tau)\right] \ldots\right]\right\rangle \, ,
\end{eqnarray}
where $H_{I}$ is the interaction Hamiltonian. As the dynamic is effectively described by a single-field model of inflation, the interactions will be the same as those derived in the context of the EFT of single-field inflation \cite{Cheung:2007st}. 

After inserting the UV cutoff of the EFT, the time integrals become of the form
\begin{eqnarray} \label{integral function}
\mathcal{F}[n, p] \equiv \left. \int_{\alpha}^0 dy \, y^n \exp \left[-p y  \right] = - p^{-1-n} \Gamma\left[n+1, p y\right] \right|_{y=\alpha}^{y=0}
\end{eqnarray} 
where $\Gamma[n,x]$ is the incomplete Gamma function, $n>-1$ and $p>0$ is a function of the different momenta. The integrand peaks at $ y_{max}= n/p$. Therefore, there are basically two distinct classes of shapes: 
\begin{itemize}
	\item (near)-collapsed shapes where $p \ll n/\alpha$:  \textit{UV sensitive};
	\item other shapes: \textit{UV and IR finite}.
\end{itemize}
The UV sensitivity is not necessarily dramatic.
If the theory behaves approximately as a weakly coupled system of massless fields for scales above the cutoff, $k/a>|m_s|$, as it is likely the case (see eq. \eqref{Lagrangian for the quadratic perturbations}), the mode functions will quickly be well described by rapidly oscillating plane waves. The oscillatory behavior will then strongly suppress the contribution of the UV to the integrals. Therefore, the results here presented for those shapes can still be trusted at the order of magnitude level. The fact that the non-Gaussianity peaks at that collapsed shapes is insensitive to this discussion.

In the next sections, I will compute the non-Gaussianities for a generic shape and then particularize the results for the equilateral ($k_i=k$) and flattened ($k_1=k_2+k_3$ for the bispectrum and $k_1=k_2+k_3+k_4$ for the trispectrum) configurations as representative shapes of those two classes.

\subsection{Bispectrum \label{sec:bispectrum}}

I start by reviewing the bispectrum computation \cite{Garcia-Saenz:2018vqf,Fumagalli:2019noh}. In Fourier space the 3-point function can generically be written as 
\begin{eqnarray}
\left\langle\zeta_{\q_1}\zeta_{\q_2}\zeta_{\q_3}  \right\rangle &=& (2\pi)^{-3/2}  \delta^{(3)}(\q_1+\q_2+\q_3) B(\q_1,\q_2,\q_3) \, , \\  \label{shape function} B(\q_1,\q_2,\q_3) &\equiv& (2 \pi)^{4} \frac{S\left(\q_{1}, \q_{2}, \q_{3}\right)}{\left(k_{1} k_{2} k_{3}\right)^{2}} P_\zeta^{2} \, \, ,
\end{eqnarray}
where $ B(\q_1,\q_2,\q_3) $ is the bispectrum and I have defined the shape function $S\left(\q_1,\q_2,\q_3\right)$ for later convenience. 
For $n=3$, the first term in eq. \eqref{in-in expansion} is
\begin{eqnarray} \label{3-point function}
\left\langle\zeta^{3}(\tau)\right\rangle &= &i \int_{-\infty}^{\tau} d \tau_{1}  \left\langle \left[H_{I,3}\left(\tau_{1}\right), \zeta^{3}(\tau) \right]\right\rangle = -2 \int_{-\infty}^{\tau} d \tau_{1}  \, \text{Im} \left[ \left\langle H_{I,3}\left(\tau_{1}\right) \zeta^{3}(\tau) \right\rangle \right] \, ,
\end{eqnarray}
where $H_{I,3}$ is the interaction Hamiltonian of cubic order given in eq. \eqref{cubic Hamiltonian}. The fact that only the imaginary part contributes is crucial to tame the exponential growth of non-Gaussianities \cite{Ferreira:2015omg,Bjorkmo:2019qno}. A similar thing happens in the 4-point function.
 
I am interested in the late time limit, $\tau \rightarrow 0$. Then, after Fourier transforming the fields using eq. \eqref{Fourier convention}, performing the Wick contractions, changing variable to $\cR$ using eq. \eqref{cR definition} and using eqs.  \eqref{H3 decomposition} and \eqref{real and imaginary part of R} the 3-point function becomes
\begin{eqnarray} \label{im re decomposition 2}
&& \left\langle\zeta_{k_1}\zeta_{k_2}\zeta_{k_3}  \right\rangle' \simeq  \nonumber \\ && \qquad-2 h \left(\frac{ \pi^2 P_{\zeta,0}}{k_1 k_2 k_3} \right)^{3/2}  \int_{-\infty}^{\tau} \frac{d \tau_{1}}{\tau_1} \, \left[  \left(\frac{2P_\zeta}{P_{\zeta,0}} \right)^{3/2} \text{Im} \left[  \frac{A}{ |c_s^2|}  \mathcal{R}^{\prime}_{k_1} \mathcal{R}^{\prime}_{k_2}\mathcal{R}^{\prime}_{k_3} + \left(\q_1 \cdot \q_2\right) \mathcal{R}_{k_1}   \mathcal{R}_{k_2} \mathcal{R}'_{k_3}  \right] \right.\nonumber \\ && \left. \qquad  - 3\left(\frac{2P_{\zeta}}{P_{\zeta,0}} \right)^{1/2} \text{Re} \left[  \frac{A}{ |c_s|^2}  \mathcal{R}^{\prime}_{k_1} \mathcal{R}^{\prime}_{k_2}\mathcal{R}^{\prime}_{k_3} + \left(\q_1 \cdot \q_2\right) \mathcal{R}_{k_1}   \mathcal{R}_{k_2} \mathcal{R}'_{k_3}  \right] \right] + \text{5 perm.} 
\end{eqnarray}
The factor of 3 in the last term comes from the 3 different options to chose the imaginary part of the external legs and the minus sign from the complex conjugation. The prime in the correlator denotes that the momentum conserving delta function was suppressed. 

Now I can evaluate the integrals using the mode functions in eq. \eqref{mode function}. The EFT is only valid for modes below the cutoff. Therefore, the time integrals are cutted off at $- k_\text{max} \tau_\text{UV} |c_s| = \alpha$ where $k_\text{max}$ is the largest momentum in the correlator\footnote{I will assume $k_1$ to be the largest momenta throughout the paper apart from sec. \ref{sec:cubic interaction} where I choose $k_4$ instead.}.
The results below agree with those found in \cite{Garcia-Saenz:2018vqf} and show that bispectrum peaks in flattened shapes and is proportional to $\alpha^3$.

\subsubsection{Terms in $\zeta'^3$}

 As mentioned before, the imaginary part ensures that the result is proportional to $e^{4x}$ thus making the 3-point function proportional to $P_\zeta^2$ in the $x \gg 1$ limit. In terms of $y=-k_1 |c_s| \tau$, the terms in $\zeta'^3$ in eq. \eqref{im re decomposition 2} simplify to
\begin{eqnarray}
&& \left\langle\zeta_{k_1}\zeta_{k_2}\zeta_{k_3}  \right\rangle'_{\zeta'^3} \simeq \\  && \qquad -(2\pi)^{5/2} \frac{P_\zeta^2}{\left(k_1 k_2 k_3\right)^{3}}  \left(\frac{1}{|c_{s}^{2}|}+1\right) \frac{3}{8|c_s|} \frac{A}{ |c_s^2|} \int_{\alpha}^{0} \frac{d y}{y} \,  \left[ \left(g^{\prime}_{k_1} f^{\prime}_{k_2} f^{\prime}_{k_3}  + \text{2 perm.} \right) +3   f^{\prime}_{k_1} f^{\prime}_{k_2} f^{\prime}_{k_3}  \right] \, ,   \nonumber 
\end{eqnarray}
where I have used eqs. \eqref{h} and \eqref{mode function}.
After evaluating the integrals and using eq. \eqref{integral function} the shape function in eq. \eqref{shape function} becomes
 \begin{eqnarray}
 S\left(k_{1}, k_{2}, k_{3}\right)_{\zeta'^3} = -\frac{3 A}{8|c_s|^3}  \left(\frac{1}{|c_{s}^{2}|}+1\right) \frac{1}{k_1 k_2 k_3}  \left[ \left[ \frac{k_2^2 k_3^2 |c_s|^3}{k_1} {\cal F}[3,p] + \text{2 perm.} \right]-3{\cal F}[3,p_t] \right] \, ,\quad \, \, 
 \end{eqnarray}
 where $p=(-k_1+k_2+k_3)/k_1$ and $p_t=(k_1+k_2+k_3)/k_1$. As anticipated in the beginning of the section, there are essentially two distinct classes of shapes. For the dominant shapes, near the flattened limit $\alpha p \ll 1 $ (c.f. fig. \ref{fig:flat3pf}), it gives
 \begin{eqnarray}
 S\left(k_{1}, k_{2}, k_{3}\right)^\text{near-flat}_{\zeta'^3} = \frac{A}{8}  \left(\frac{1}{|c_{s}^{2}|}+1\right)   \frac{k_2 k_3 }{ k_1^2} \alpha^3 \left[1- \frac{3}{4}p \alpha +{\cal O}(p \alpha)^2\right] \, .
 \end{eqnarray}
 Note that the 2 permutations did not contribute to leading order in $\alpha$. For example, for $k_1/2=k_2=k_3=k$ it yields
  \begin{eqnarray} \label{shape function zeta'3}
 S\left(2k, k, k\right)_{\zeta'^3} \simeq \frac{A}{32}  \left(\frac{1}{|c_{s}^{2}|}+1\right)   \alpha^3 \, .
 \end{eqnarray}
For the remaining shapes the result converges in the $\alpha \gg 1 $ limit yielding
 \begin{eqnarray}
 S\left(k_{1}, k_{2}, k_{3}\right)^\text{non-flat}_{\zeta'^3} \simeq -\frac{3}{4}A  \left(\frac{1}{|c_{s}^{2}|}+1\right)  \frac{1}{k_1 k_2 k_3} \left[ -\left(\frac{k_2^2  k_3^2}{k_1} \frac{1}{p^3}+ \text{2 perm.} \right) +\frac{3}{p_t^3} \right]  \, .
 \end{eqnarray}
In particular, in the equilateral limit $k_1=k_2=k_3$ it gives
\begin{eqnarray}
S\left(k,k,k\right)_{\zeta'^3} \simeq  \frac{13}{6} A  \left(\frac{1}{|c_{s}^{2}|}+1\right) \, .
\end{eqnarray}

\subsubsection{Terms in $\zeta' (\partial \zeta)^2$}
Regarding the terms in $\zeta' (\partial \zeta)^2$ in eq. \eqref{im re decomposition 2}, after using eq. \eqref{mode function}, I get that
\begin{eqnarray}
\left\langle\zeta_{k_1}\zeta_{k_2}\zeta_{k_3}  \right\rangle'_{\zeta' (\partial \zeta)^2} &\simeq &  -(2\pi)^{5/2} \frac{P_\zeta^2}{\left(k_1 k_2 k_3\right)^{3}}  \left(\frac{1}{|c_{s}^{2}|}+1\right) \frac{\left(\q_1 \cdot \q_2\right)}{16|c_s|} \left[ \int_{\alpha}^{0} \frac{d y}{y} \,  \left[  g_{k_1}   f_{k_2} f'_{k_3} + \text{2 perm.} \right] + \right.  \nonumber \\ &&  + \left. 3   f_{k_1}   f_{k_2} f'_{k_3}  \right] + \text{5 perm.} 
\end{eqnarray}
The shape function is in this case given by
\begin{eqnarray}
&& S\left(k_{1}, k_{2}, k_{3}\right)_{\zeta' (\partial \zeta)^2} = \nonumber \\ && \qquad -\frac{\left(\q_1 \cdot \q_2\right)}{16 } \left(\frac{1}{|c_{s}^{2}|}+1\right) \frac{1 }{k_1 k_2 k_3} \left[ \left( \frac{ k_3^2}{k_1^2}  \left( k_1 {\cal F}[1,p] +(k_2-k_1) {\cal F}[2,p] -k_2 {\cal F}[3,p] \right) \right. \right. \nonumber \\ && \left. \left. \qquad  + \text{2 perm.} \right)  - 3\frac{ k_3^2}{k_1^2}  \left(k_2 {\cal F}[3,p_t]+(k_2+k_1) {\cal F}[2,p_t] +k_1 {\cal F}[1,p]  \right) \right] +\text{5 perm.} 
\end{eqnarray}
For near-flattened shapes, $p \alpha \ll 1$, in the large-$\alpha$ limit it simplifies to
\begin{eqnarray}
S\left(k_{1}, k_{2}, k_{3}\right)_{\zeta' (\partial \zeta)^2}^\text{near-flat} &\simeq& - \frac{ \left(\q_1 \cdot \q_2\right)}{48}  \left(\frac{1}{|c_{s}^{2}|}+1\right)\alpha^3 \frac{ k_3 }{k_1^3}   \left[1- \frac{3}{4}p \alpha +{\cal O}(p \alpha)^2\right] + \text{5 perm.} \,\,
\end{eqnarray}
In such configuration $\q_1 \cdot \q_2 =-k_1 k_2$ and  $\q_2 \cdot \q_3 =k_2 k_3 $ which implies that
\begin{eqnarray} \label{shape function zeta'zeta^2}
S\left(2k, k, k\right)_{\zeta' (\partial \zeta)^2} &\simeq&   \left(\frac{1}{|c_{s}^{2}|}+1\right)  \frac{ \alpha^3}{32} \,.
\end{eqnarray}
For other shapes, it gives instead
\begin{eqnarray}
 S\left(k_{1}, k_{2}, k_{3}\right)_{\zeta' (\partial \zeta)^2}^\text{non-flat} &=&  \frac{\left(\q_1 \cdot \q_2\right)}{16}  \left(\frac{1}{|c_{s}^{2}|}+1\right) \frac{1}{k_1 k_2 k_3}   \left[ \left(\frac{k_3^2 }{k_1^2 p^3}  \left(k_1 (p-1) p+k_2 (p-2)\right) \right. \right.  \nonumber \\ &&  \left.  \left. + \text{ 2 perm.}\right) -3 \frac{k_3^2  }{k_1^2 p_t^3} \left(k_1 p_t \left(p_t+1\right)+k_2 \left(p_t+2\right)\right)\right] +\text{5 perm.} 
\end{eqnarray}
which, in the equilateral limit, reduces to
\begin{eqnarray}
S\left(k, k, k\right)&=&- \frac{5}{24}  \left(\frac{1}{|c_{s}^{2}|}+1\right)\,.
\end{eqnarray}

\subsection{Trispectrum \label{sec:trispectrum}}
\begin{figure} 
	\centering
	\begin{tikzpicture}[baseline={(current bounding box.center)}]
	\begin{feynman}
	\vertex (a) ;
	\vertex [above  left=of a] (b);
	\vertex [below  left=of a] (d) ;
	\vertex [      right=of a] (f) ;
	\vertex [above right=of f] (c) ;
	\vertex [below right=of f] (e);
	\diagram* {
		(b) --  (a) -- (d),
		(a) --  (f),
		(c) -- (f) --  (e),
	};
	\vertex [left=0.5em of a] {\(H_{I,3}\)};
	\vertex [right=0.5em of f] {\(H_{I,3}\)};
	\end{feynman} 
	\end{tikzpicture} \qquad \qquad
	\begin{tikzpicture}[baseline={(current bounding box.center)}]
	\begin{feynman}
	\vertex (a) ;
	\vertex [above  left=of a] (b);
	\vertex [below  left=of a] (d) ;
	\vertex [above right=of a] (c) ;
	\vertex [below right=of a] (e);
	\diagram* {
		(b) --  (a) -- (d),
		(c) --  (a) -- (e),
	};
	\vertex [right=0.5em of a] {\(H_{I,4}\)};
	\end{feynman} 
	\end{tikzpicture}
	\caption{Contributions to the 4-point function: exchange diagram and contact interaction. \label{fig:4ptX}}
\end{figure}
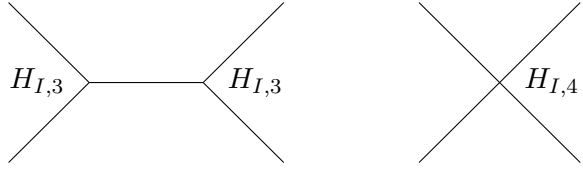
The next step is to compute the contribution to the trispectrum from the two diagrams in fig. \ref{fig:4ptX}. 
The results will share many similarities with the bispectrum calculation: they peak in shapes where all the momenta are (anti-)collinear, which in the case of the exchange diagram does not correspond solely to the $k_1 \rightarrow k_2+k_3+k_4$ limit, and the remaining shapes are UV and IR convergent in the $\alpha \gg 1$ limit. Moreover, I find that the contribution from the leading flattened shapes scale as $\alpha^6$ for the exchange diagram, in agreement with \cite{Fumagalli:2019noh}, and so dominates over the contact interaction where they scale as $\alpha^5$. This can be easily understood from the time integrals in the in-in formalism which in the flattened limit read $\left( \int^{\alpha}_0 y^2 dy \right)^2 $ for the exchange diagram and $\int^{\alpha}_0 y^4 dy $ for the contact term .

The two diagrams in fig. \ref{fig:4ptX} correspond to the first two corrections in the in-in expansion and are given by
\begin{eqnarray} \label{first two terms in the in-in formalism}
\left\langle\zeta^{4}(\tau)\right\rangle_\text{conn.} &\simeq&  i \int_{-\infty}^{\tau} d \tau_{1}  \left\langle \left[H_{I,4}\left(\tau_{1}\right), \zeta^{4}(\tau) \right]\right\rangle  - \\ && -\int_{-\infty}^{\tau} d \tau_{1} \int_{-\infty}^{\tau_{1}} d \tau_{2}\left\langle\left[H_{I,3}\left(\tau_{2}\right),\left[H_{I,3}\left(\tau_{1}\right), \zeta^{4}(\tau)\right]\right]\right\rangle \equiv  \left\langle\zeta^{4}(\tau)\right\rangle_{c,1} + \left\langle\zeta^{4}(\tau)\right\rangle_{c,2}  \nonumber
\end{eqnarray}
where $H_{I,4}$ is the interaction Hamiltonian of quartic order given in eq. \eqref{quartic Hamiltonian}.
In Fourier space the 4-point function is generically given by
\begin{eqnarray} \label{4pf}
\left\langle\zeta_{k_1}\zeta_{k_2}\zeta_{k_3}\zeta_{k_4}\right\rangle_\text{conn.}=  \frac{(2 \pi)^{3} P_{\zeta}^{3}}{(k_{1} k_2 k_3 k_4)^{9/4}} \mathcal{T}\left(k_{1}, k_{2}, k_{3}, k_{4} \right)\,  \delta^{(3)} \left(\sum_{\mathbf{i}} \mathbf{k}_{\mathbf{i}}\right) \, .
\end{eqnarray}
The goal now is to calculate $\cT$ from the different contributions. 

\subsubsection{Quartic interaction}

We start by computing $\left\langle\zeta^{4}(\tau)\right\rangle_{c,1}$, the contact interaction. Using eqs. \eqref{Fourier convention} and \eqref{cR definition} the correlator becomes
\begin{eqnarray}
\left\langle\zeta_{k_1}\zeta_{k_2}\zeta_{k_3}\zeta_{k_4}\right\rangle_{c,1} =-2 (\pi^2 P_{\zeta,0})^2 \int_{-\infty}^{\tau} d \tau_{1}  \text{Im} \left[\left\langle H_{I,4}\left(\tau_{1}\right), \left(\cR_{k_1}\cR_{k_2}\cR_{k_3}\cR_{k_4}\right)_\tau \right\rangle  \right] \, . \label{first correction}
\end{eqnarray}
Just like in the bispectrum, the imaginary part is crucial to tame exponential growth of non-Gaussianities. Now I perform the Wick contractions and separate the integrand in two terms depending on whether I take the real or imaginary part of the external legs. After using eqs. \eqref{real and imaginary part of R}, \eqref{4pf} and inserting the interaction Hamiltonian in eq. \eqref{tilde H4} the trispectrum becomes
\begin{eqnarray}
\mathcal{T}\left(k_{1}, k_{2}, k_{3}, k_{4}\right) &= &  \frac{1}{32 |c_s|}  \left(\frac{1}{|c_{s}^{2}|}+1\right)  \frac{1}{\left(k_1 k_2 k_3 k_4\right)^{3/4}}   \sum_{i=1}^3 \mathcal{I}_i(k_1,k_2,k_3,k_4) \, ,
\end{eqnarray}
where 
\begin{eqnarray}  \label{I1}
\mathcal{I}_1(k_1,k_2,k_3,k_4) & \simeq & 24 B  \int_{\alpha}^{0}  \frac{dy}{-k_1 |c_s|} \left[ -\left[ \, g^{\prime}_{k_1} f^{\prime}_{k_2}f^{\prime}_{k_3} f^{\prime}_{k_4}  + \text{3 perm.}  \right] -f^{\prime}_{k_1} f^{\prime}_{k_2}f^{\prime}_{k_3} f^{\prime}_{k_4} \right]  \, , \, \\ \label{I2}
\mathcal{I}_2 (k_1,k_2,k_3,k_4) & \simeq & - 4C (\q_1 \cdot \q_2)  \int_{\alpha}^{0}  \frac{dy}{-k_1 |c_s|} \left[- \left[  g_{k_1} f_{k_2} f^{\prime}_{k_3} f^{\prime}_{k_4} + \text{3 perm.}  \right]-   \nonumber  \right. \\ && \left. -f_{k_1} f_{k_2}f^{\prime}_{k_3} f^{\prime}_{k_4} \right]+ \text{5 perm.} \, ,  \\ \label{I3}
\mathcal{I}_3 (k_1,k_2,k_3,k_4) & \simeq & |c_s^2| \left( \q_1 \cdot \q_2 \right) \left(\q_3 \cdot \q_4 \right)  \int_{\alpha}^{0}  \frac{dy}{-k_1 |c_s|} \left[- \left[   g_{k_1} f_{k_2} f_{k_3} f_{k_4} + \text{3 perm.}  \right] - \right. \nonumber  \\ && \left.  -f_{k_1} f_{k_2} f_{k_3} f_{k_4}  \right] + \text{5 perm.}
\end{eqnarray}
In the last step, I inserted the mode functions in eq. \eqref{mode function}, changed the time variable to $y=-k_1 \tau |c_s|$ and inserted the EFT cutoff in the time integral assuming $k_1$ to be the largest momenta. 

\paragraph{Terms in $\zeta'^4$:}

Using eqs. \eqref{integral function} and \eqref{mode function 2} the integral in eq. \eqref{I1} simplifies to
\begin{eqnarray}
\mathcal{I}_1 (k_1,k_2,k_3,k_4) = - 24 B |c_s^3|  \frac{\left(k_2 k_3 k_4  \right)^2}{k_1^3 } \left[  {\cal F} \left[4, q y \right]  + \text{3 perm.} -  {\cal F} \left[4, q_t y \right] \right]
\end{eqnarray} 
where $q \equiv  (-k_1+k_2+k_3+k_4)/k_1 $ and $q_t=(k_1+k_2+k_3+k_4)/k_1$. The 3 permutations are similar, up to the appropriate redefinition of $q$. However, they do not contribute to the flattened limit, $q \ll 1/\alpha$, to leading order in $\alpha$ which reads
\begin{eqnarray}
\mathcal{I}_1^\text{near-flat}(k_1,k_2,k_3,k_4) &= & 24B \alpha^5 |c_s|^3  \frac{\left(k_2 k_3 k_4  \right)^2 }{k_1^3}   \left(\frac{1}{5} - \frac 1 6 q \alpha  + {\cal O}(q \alpha )^2 \right) \, ,
\end{eqnarray}
and for $k_1/3=k_2=k_3=k_4=k$ yields the trispectrum
\begin{eqnarray}
\mathcal{T}_{1}(3k,k,k,k) = \frac{B   |c_s|^2}{180 \cdot 3^{3/4}}  \left(\frac{1}{|c_{s}^{2}|}+1\right) \alpha^5 \, . \label{T zeta'4}
\end{eqnarray}
The remaining shapes are UV and IR convergent in the $\alpha \gg 1$ limit and given by 
\begin{eqnarray} 
\mathcal{I}_1^\text{non-flat}(k_1,k_2,k_3,k_4) &= & 24^2 B |c_s|^3 \left[ \frac{\left(k_2 k_3 k_4  \right)^2}{k_1^3 q^5}   + \text{3 perm.} - \frac{\left(k_2 k_3 k_4  \right)^2}{k_1^3 q_t^5}  \right] \, .
\end{eqnarray}
In particular, for equilateral shapes
\begin{eqnarray} 
\mathcal{T}_1 (k,k,k,k)&= & \frac{855}{512} B |c_s|^2 \, .
\end{eqnarray}

\paragraph{Terms in $\zeta'^2 (\partial \zeta)^2$:}

Following steps similar to those of above,  $\mathcal{I}_2$ in eq. \eqref{I2} evaluates to 
\begin{eqnarray}
\mathcal{I}_2 (k_1,k_2,k_3,k_4) &=&   -4C |c_s|   (\q_1 \cdot \q_2) \frac{\left(k_3 k_4\right)^2 }{k_1^4}   \times  \\ &&  \left[  \left[ k_2 {\cal F}\left[4, q y\right] +(k_1-k_2) {\cal F}\left[3, q y\right]  - k_1 {\cal F}\left[2, q y\right] + \text{3 perm.} \right] \right.   \nonumber \\ &&  \left. + \left[ k_2 {\cal F}\left[4, q_t y\right] +(k_1+k_2) {\cal F}\left[3, q_t y\right]  + k_1 {\cal F}\left[2, q_t y\right] \right] \right] + \text{5 perm.}  \nonumber
\end{eqnarray}
In the flattened limit, $q \ll 1/\alpha $, the leading terms are
\begin{eqnarray}
&&\mathcal{I}_2^\text{near-flat}(k_1,k_2,k_3,k_4) = 4 C |c_s| \times \\&& \quad \quad    \left[ \left(\q_1\cdot \q_2\right)  \frac{k_3^2 k_4^2}{60 k_1^4}  \alpha^3 \left(5 (k_3+k_4) \left(3 \alpha-4\right)+4 k_2 \left(3 \alpha^2-5\right)\right) + \text{5 perm.} \right] +{\cal O}\left(q \alpha \right)  \nonumber \,.
\end{eqnarray}
When $q \rightarrow 0$ all momenta are collinear apart from $k_1$ which is anti-collinear. Therefore, for $\alpha \gg 1$ the trispectrum is given by
\begin{eqnarray}
 \mathcal{T}_2(3k,k,k,k)   &=& -  \frac{C }{180  \cdot 3^{3/4}}  \left(\frac{1}{|c_{s}^{2}|}+1\right)  \alpha^5  \,. \label{T zeta'2 dzeta2}
 \end{eqnarray}
For other shapes, in the $\alpha \gg 1 $ limit, I get instead
\begin{eqnarray}
\mathcal{I}_2 ^\text{non-flat}(k_1,k_2,k_3,k_4)  &=&  -8 \left(\q_1\cdot \q_2\right) C  |c_s| \frac{ k_3^2 k_4^2 }{k_1^4} \left[ \frac{1 }{q^5} \left( \left(k_1 (q-3) q+3 k_2 (q-4)\right)+ \text{3 perm.} \right)- \right. \nonumber \\ && \left. \frac{1 }{q_t^5} \left(k_1 q_t \left(q_t+3\right)+3 k_2 \left(q_t+4\right)\right)\right]  + \text{5 perm.}
\end{eqnarray}

\paragraph{Terms in $(\partial \zeta)^4$:}

Finally I look at ${\cal I}_3$ in eq. \eqref{I3}. After inserting the mode functions in eq. \eqref{mode function} it reads 
\begin{eqnarray} \label{I3 int}
&& \mathcal{I}_3(k_1,k_2,k_3,k_4) =  |c_s| \frac{ \left( \q_1 \cdot \q_2 \right) \left(\q_3 \cdot \q_4 \right)}{k_1^4}  \int_{\alpha}^{0}  dy \left[ (y-1) \left(k_2 y+k_1\right) \left(k_3 y+k_1\right) \left(k_4 y+k_1\right) e^{-q y}   \right. \nonumber \\ && \left.  \qquad \qquad +\text{3 perm.} +\frac{1}{k_1^4 |c_s|} (y+1)  \left(k_2 y+k_1\right) \left(k_3 y+k_1\right) \left(k_4 y+k_1\right) e^{-y q_t} \right] + \text{5 perm.} \, 
\end{eqnarray}
The expression in terms of the function ${\cal F}$ defined in eq. \eqref{integral function} is long and not very informative so I give it in eq. \eqref{I3 app}. 
In the near-flattened limit, and for $\alpha \gg 1$, it gives
\begin{eqnarray}
\mathcal{I}_3^\text{near-flat}(k_1,k_2,k_3,k_4)  & \simeq &  - \frac  1 5 |c_s| \alpha^5  \frac{k_2 k_3 k_4}{k_1^4} \left[ \left( \q_1 \cdot \q_2 \right) \left(\q_3 \cdot \q_4 \right)   + 5 \text{ perm.} \right]  + \cO(q\alpha)\, , \,  \, 
\end{eqnarray}
and yields the trispectrum
\begin{eqnarray}
\mathcal{T}_3(3k,k,k,k)    &=&  \frac{1}{720 \cdot 3^{3/4} }  \left(\frac{1}{|c_{s}^{2}|}+1\right)  \alpha^5\,.  \label{T dzeta4}
\end{eqnarray}
For other shapes I get instead the result in eq. \eqref{I3 nonflat}. 

To sum up, the contributions to the trispectrum from the contact interaction peak in the flattened shape, proportionally to $\alpha^5$, while other shapes are $\alpha$-independent in the large-$\alpha$ limit. 

\subsubsection{Cubic interaction \label{sec:cubic interaction}}

Finally, I pass to the main contribution to the trispectrum, the term $\left< \zeta^4 (\tau)\right>_{c,2}$ in eq. \eqref{first two terms in the in-in formalism} corresponding to the exchange diagram in fig. \ref{fig:4ptX}. 
I start by expanding the integrand into
\begin{eqnarray}
\left\langle\left[\hat{H}_I\left(\tau_{2}\right),\left[\hat{H}_I\left(\tau_{1}\right), \hat{\zeta}^{4}(\tau)\right]\right]\right\rangle = 2 \operatorname{Re} \left\langle\hat{H}_I\left(\tau_{2}\right) \hat{H}_I\left(\tau_{1}\right) \hat{\zeta}^{4}(\tau)  -\hat{H}_I\left(\tau_{2}\right) \hat{\zeta}^{4}(\tau) \hat{H}_I\left(\tau_{1}\right)\right\rangle  \,. \quad
\end{eqnarray}
Then, after changing from $\zeta$ to $\cR$ using eq. \eqref{cR definition} and inserting the cubic Hamiltonian in eq. \eqref{H3 decomposition} the 4-point function is given, in Fourier space, by
\begin{eqnarray}
&& \left\langle\zeta_{k_1}\zeta_{k_2}\zeta_{k_3}\zeta_{k_4}\right\rangle_{c,2}  = - 2 h^2 \left( \pi^2 P_{\zeta,0} \right)^2 \int_{-\infty}^{\tau} d \tau_{1} \int_{-\infty}^{\tau_{1}} d \tau_{2}   \nonumber \\ && \qquad  \operatorname{Re} \left\langle \tilde{H}_{I,3}\left(\tau_{2} \right) \left( \tilde{H}_{I,3}\left(\tau_{1} \right)\left(\cR_{k_1}\cR_{k_2}\cR_{k_3}\cR_{k_4} \right)_\tau- \left(\cR_{k_1}\cR_{k_2}\cR_{k_3}\cR_{k_4} \right)_\tau \tilde{H}_{I,3}\left(\tau_{1} \right) \right) \right\rangle \, . \quad
\end{eqnarray}
Each $\tilde{H}_{I,3}$ contains two distinct terms so there are 4 different combinations. I define the functions ${\cal K}_i$ associated with each of those terms as
\begin{eqnarray} \label{cubic terms as a function of K}
\left\langle\zeta_{k_1}\zeta_{k_2}\zeta_{k_3}\zeta_{k_4}\right\rangle_{c,2}  =-\frac{\left(\frac{1}{|c_{s}^{2}|}+1\right)^2}{2^9 |c_s|^2} \frac{(2 \pi)^{3} P_{\zeta}^{3} }{\left(k_1 k_2 k_3 k_4 \right)^{3}} \delta^{(3)} \left( \sum_{i=1}^4 k_i \right) \sum_{i=1}^4  {\cal K}_i  
\end{eqnarray}
where $P_\zeta^3/2$ was factored out for convenience. The trispectrum in eq. \eqref{4pf} is then related to $\cK_i$ by
\begin{eqnarray} \label{trispectrum coefficient}
\mathcal{T}\left(k_{1}, k_{2}, k_{3}, k_{4}\right) &= &  -\left(\frac{1}{|c_{s}^{2}|}+1\right)^2 \frac{1}{2^9 |c_s|^2}   \frac{1}{ \left(k_1 k_2 k_3 k_4\right)^{3/4}}  \sum_{i=1}^4 \mathcal{K}_i(k_1,k_2,k_3,k_4) \, .
\end{eqnarray}
The function ${\cal K}_1$ is associated with two insertions of $\tilde{H}_{3,2}$, defined in eq. \eqref{H3 decomposition}, and it is given by
\begin{eqnarray}
&& {\cal K}_1 \left(k_{1}, k_{2}, k_{3}, k_{4}\right)=  - 2\times 2 \times (3\times 3\times 4)\left( \frac{A}{ |c_s^2|} \right)^2\left(k_1 k_2 k_3 k_4 \right)^{3} e^{-6x} \int_{-\infty}^{\tau} \frac{d \tau_{1}}{\tau_1} \int_{-\infty}^{\tau_{1}} \frac{d \tau_{2}}{\tau_2} \nonumber   \\ &&  \quad \text{Im}  \left[ \left(\mathcal{R}^{\prime}_{k_1} \mathcal{R}^{\prime}_{k_2}\mathcal{R}^{\prime}_{k_{12}}\right)_{\tau_2}  \left(\mathcal{R}_{k_1} \mathcal{R}_{k_2}\right)^*_{\tau}  \mathcal{R}^{\prime\,*}_{k_{12}, \tau_1} \right]  \text{Im} \left[ \left(\mathcal{R}^{\prime}_{k_4} \mathcal{R}^{\prime}_{k_3} \right)_{\tau 1} \left(\mathcal{R}_{k_3} \mathcal{R}_{k_4}\right)^*_{\tau} \right]  + \text{5 perm.} 
\end{eqnarray}
where $\q_{12} = \q_1+\q_2$. Some of the pre-factors are shared by the several $\cK_i$ so I explain here their origin. The factor $ 3\times 3\times 4$ comes from permutations within the first vertex times permutations within the second vertex and permutations $1 \leftrightarrow 2, 3 \leftrightarrow 4$ in the external legs\footnote{The previous expression, jointly with eq. \eqref{trispectrum coefficient} agrees with eq. (21) of \cite{Fumagalli:2019noh}.}. The two factors of $2$ come from picking one of the imaginary parts and to compensate the factor of $1/2$ which I factored out in eq. \eqref{cubic terms as a function of K}.
 The $\cK_2$ term is associated with one insertion of each vertex, $\tilde{H}_{3,1}$ and $\tilde{H}_{3,2}$, and is given by
\begin{eqnarray} \label{K2}
&& {\cal K}_{2}\left(k_{1}, k_{2}, k_{3}, k_{4}\right) 
= -4 \times \left(3 \times 4 \right) \left( \frac{A}{ |c_s^2|} \right) \left(k_1 k_2 k_3 k_4 \right)^{3} e^{-6x} \int_{-\infty}^{\tau} \frac{d \tau_{1}}{\tau_1} \int_{-\infty}^{\tau_{1}} \frac{d \tau_{2}}{\tau_2}    \nonumber \\ && \quad \text{Im}\left[   \left[  \left(\q_1 \cdot \q_2\right)   \left( \mathcal{R}^{\prime}_{k_{12}} \mathcal{R}_{k_1} \mathcal{R}_{k_2}  \right)_{\tau_2}  \mathcal{R}^{\prime\, *}_{k_{12}, \tau_1}  \left(\mathcal{R}_{k_1} \mathcal{R}_{k_2}\right)^*_{\tau}  \right] \text{Im} \left[ \left(\mathcal{R}^{\prime}_{k_4} \mathcal{R}^{\prime}_{k_3} \right)_{\tau_1} \left(\mathcal{R}_{k_3} \mathcal{R}_{k_4}\right)^*_{\tau} \right] + \right. \nonumber \\ && \quad \left. + \text{2 perm.}  \right] + \text{5 perm.} \, .
\end{eqnarray}
 ${\cal K}_3$ is similar to ${\cal K}_2$ but with the vertices in the opposite order
\begin{eqnarray} \label{K3}
&& {\cal K}_3 \left(k_{1}, k_{2}, k_{3}, k_{4}\right)
= - 4 \times \left(3 \times 4 \right) \left( \frac{A}{ |c_s^2|} \right) \left(k_1 k_2 k_3 k_4 \right)^{3} e^{-6x} \int_{-\infty}^{\tau} \frac{d \tau_{1}}{\tau_1} \int_{-\infty}^{\tau_{1}} \frac{d \tau_{2}}{\tau_2}    \nonumber \\ &&\quad   \text{Im}  \left[   \left( \mathcal{R}^{\prime}_{k_{12}}  \mathcal{R}'_{k_3} \mathcal{R}'_{k_4}  \right)_{\tau_2}  \mathcal{R}^{\prime\, *}_{k_{12}, \tau_1}  \left(\mathcal{R}_{k_3} \mathcal{R}_{k_4}\right)^*_{\tau}  \right] \text{Im} \left[\left(\q_1 \cdot \q_2\right)   \left(\mathcal{R}_{k_1} \mathcal{R}_{k_2} \right)_{\tau_1} \left(\mathcal{R}_{k_1} \mathcal{R}_{k_2}\right)^*_{\tau} + \right. \nonumber \\ && \quad  \left. + \text{2 perm.}   \right]  + \text{5 perm.}
\end{eqnarray}
Finally, $\cK_4$ is associated with two insertions of $\tilde{H}_{3,1}$ and given by
\begin{eqnarray} \label{K4}
&& {\cal K}_4 \left(k_{1}, k_{2}, k_{3}, k_{4}\right)
= -4 \times  4 \left(k_1 k_2 k_3 k_4 \right)^{3} e^{-6x} \int_{-\infty}^{\tau} \frac{d \tau_{1}}{\tau_1} \int_{-\infty}^{\tau_ {1}} \frac{d \tau_{2}}{\tau_2}   \nonumber \\ && \quad  \text{Im}  \left[\left[  \left(\q_1 \cdot \q_2 \right)  \left( \mathcal{R}^{\prime}_{k_{12}, \tau_2} \mathcal{R}_{k_1} \mathcal{R}_{k_2}  \right)_{\tau_2} \mathcal{R}^{\prime\, *}_{k_{12}, \tau_1}   \left(\mathcal{R}_{k_1} \mathcal{R}_{k_2}\right)^*_{\tau}  \right] \text{Im} \left[\left(\q_3 \cdot \q_4\right)   \left(\mathcal{R}_{k_4} \mathcal{R}_{k_3} \right)_{\tau_1} \left(\mathcal{R}_{k_3} \mathcal{R}_{k_4}\right)^*_{\tau}  \right.  \right. \nonumber \\ &&  \left. \left. \quad + \text{2 perm.}   \right]  + \text{2 perm.}  \right]+ \text{5 perm.}
\end{eqnarray}
Similarly to the previous sections, the time integrals are of the form \begin{eqnarray}
\int d\tau_1 \tau_1^n e^{\left( k_1 + k_2 -k_{12}\right) \tau_1}  \int d \tau_2 \tau_2^m e^{\left( k_3 + k_4 - k_{12}\right) \tau_2} \,
\end{eqnarray}
and become UV dominated when the exponent becomes smaller than $1/\alpha$. However, this time such shapes are not only those in the flattened limit, $k_4 \rightarrow k_1+k_2+k_3$, but also other collapsed shapes where all momenta are still (anti-)collinear. Nevertheless, as all such shapes give similar results I will again choose the flattened limit as representative.

In the previous sections, I have assumed $k_1$ to be the largest momenta and introduce the UV cutoff in the variable $y=-k_1 \tau |c_s|$. For convenience, I assume in this subsection $k_4$ to be the largest momenta and define instead $z=-k_4 \tau |c_s|$. 

\paragraph{Terms in ${\cal K}_1 $:}
\begin{figure}
	\centering
	\includegraphics[width=0.48\linewidth]{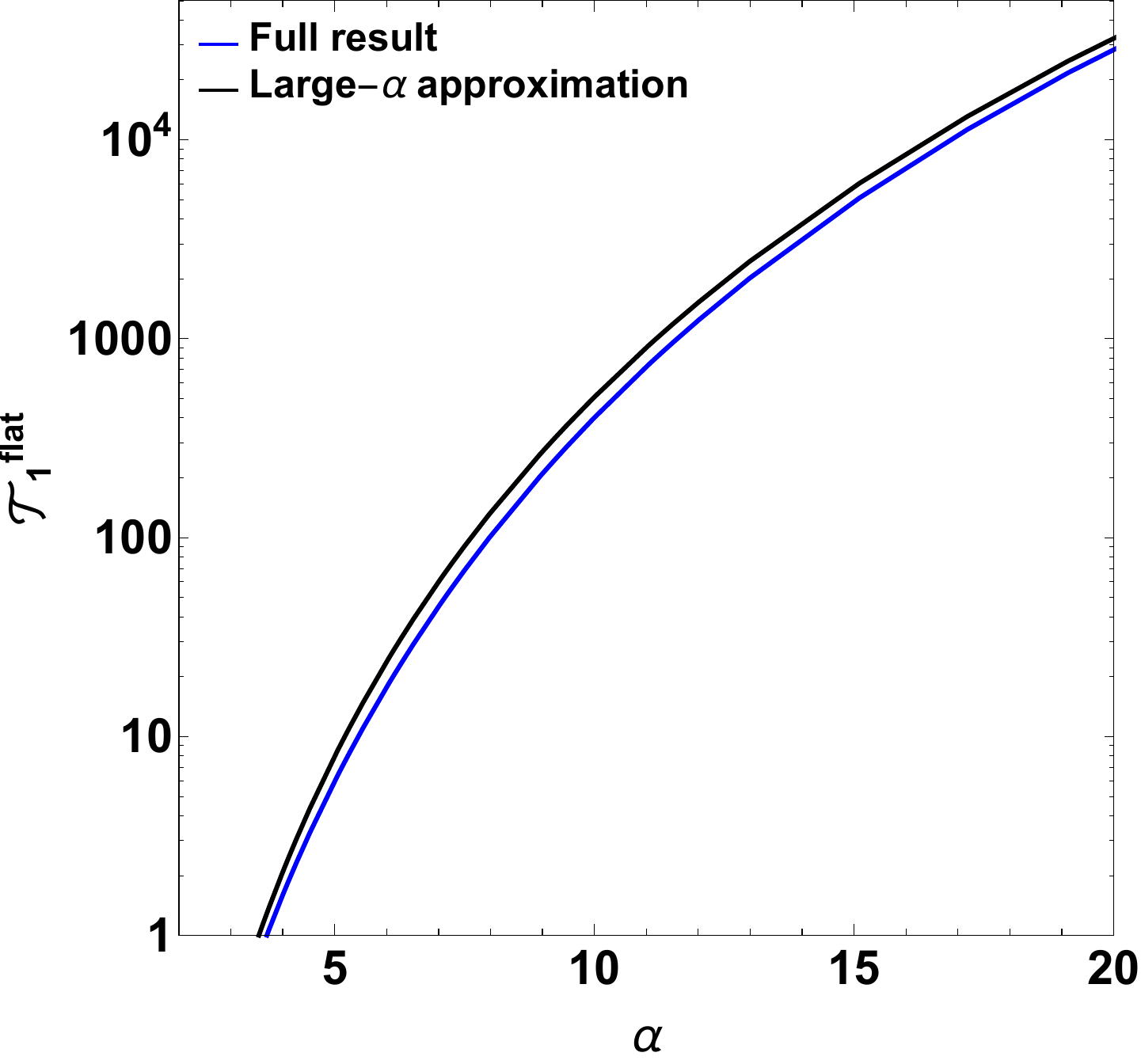}
	\caption{Contribution of $\cK_1$ to the trispectrum, in the flattened shape and as a function of $\alpha$. I plot the full expression and compare it with the large-$\alpha$ approximation.}
	\label{fig:T1flat}
\end{figure}
Let me start by the terms in ${\cal K}_1$ which are the most symmetric. After inserting the mode functions in eqs. \eqref{mode function} and \eqref{mode function 2} I get in the large $x$ limit
\begin{eqnarray}
&& {\cal K}_1 \left(k_{1}, k_{2}, k_{3}, k_{4}\right)
=  -144 \left( \frac{A}{ |c_s^2|} \right)^2 \frac{k_2^2 k_3^2 k_1^2 k_{12} |c_s|^6}{k_4^4}  \int_{\alpha}^{0} \frac{d z_1}{z_1}  \int_{\alpha}^{z_1} \frac{d z_2}{z_2}   z_1^3 z_2^3  \times  \\ && \quad  \quad  \left(e^{\frac{2 k_3 z_1}{k_4}}+e^{2 z_1}-2\right) \left(e^{\frac{2 k_1 z_2}{k_4}}+e^{\frac{2 k_2 z_2}{k_4}}+e^{\frac{2 k_{12} z_2}{k_4}}-e^{\frac{2 k_{12} z_1}{k_4}}-2\right) e^{-(p_1 z_2+p_2 z_1)} + \text{5 perm.}
\nonumber
\end{eqnarray}
where $p_1 \equiv (k_1+k_2+k_{12})/k_4$ and $p_2 \equiv (k_3+k_4+k_{12})/k_4$. The leading shapes are those such that $p_1, p_2 \rightarrow 0$ simultaneously. As mentioned above, such shapes correspond to limits where the quadrilateral formed by the 4 momenta collapses. I look at the shape $k_1=k_2=k_3=k_4/3=k$ as representative. In that case, $k_{12}=2 k$ (fig. \ref{fig:flat3pf}) and in large-$\alpha$ limit it gives
\begin{eqnarray}
{\cal K}_{1}^\text{flat} \left(k, k,k, 3k\right)
=-  3 \times \frac{144}{729}  \left( \frac{A}{ |c_s^2|} \right)^2  \alpha^6 |c_s|^{6}   k^3 \, ,
\end{eqnarray}
where the last factor of three comes from the fact that only the permutations of the external legs where $\cR_4$ is in the innermost commutator contribute to the leading order in $\alpha$. Plugging the factors in eq. \eqref{trispectrum coefficient} the trispectrum becomes 
\begin{eqnarray} \label{Trispectrum zeta'3}
{\cal T}_1^\text{flat}  \left(k, k,k, 3k\right)&=& -  \frac{1}{ 2^9 3^{3/4} |c_s|^2 k^3} \left(\frac{1}{|c_{s}^{2}|}+1\right)^2  {\cal K}_1^\text{flat} \left(k, k,k, 3k\right) \nonumber \\ &\simeq& \frac{A^2}{864 \cdot 3^{3/4}}   \alpha^6  \left(\frac{1}{|c_{s}^{2}|}+1\right)^2 \, .
\end{eqnarray}
This result is $3$ times larger than the result obtained in  \cite{Fumagalli:2019noh}.
In fig. \ref{fig:T1flat} I show that the large-$\alpha$ expression is indeed a good approximation for the full result. 

\begin{figure}
	\centering
	\includegraphics[width=0.48\linewidth]{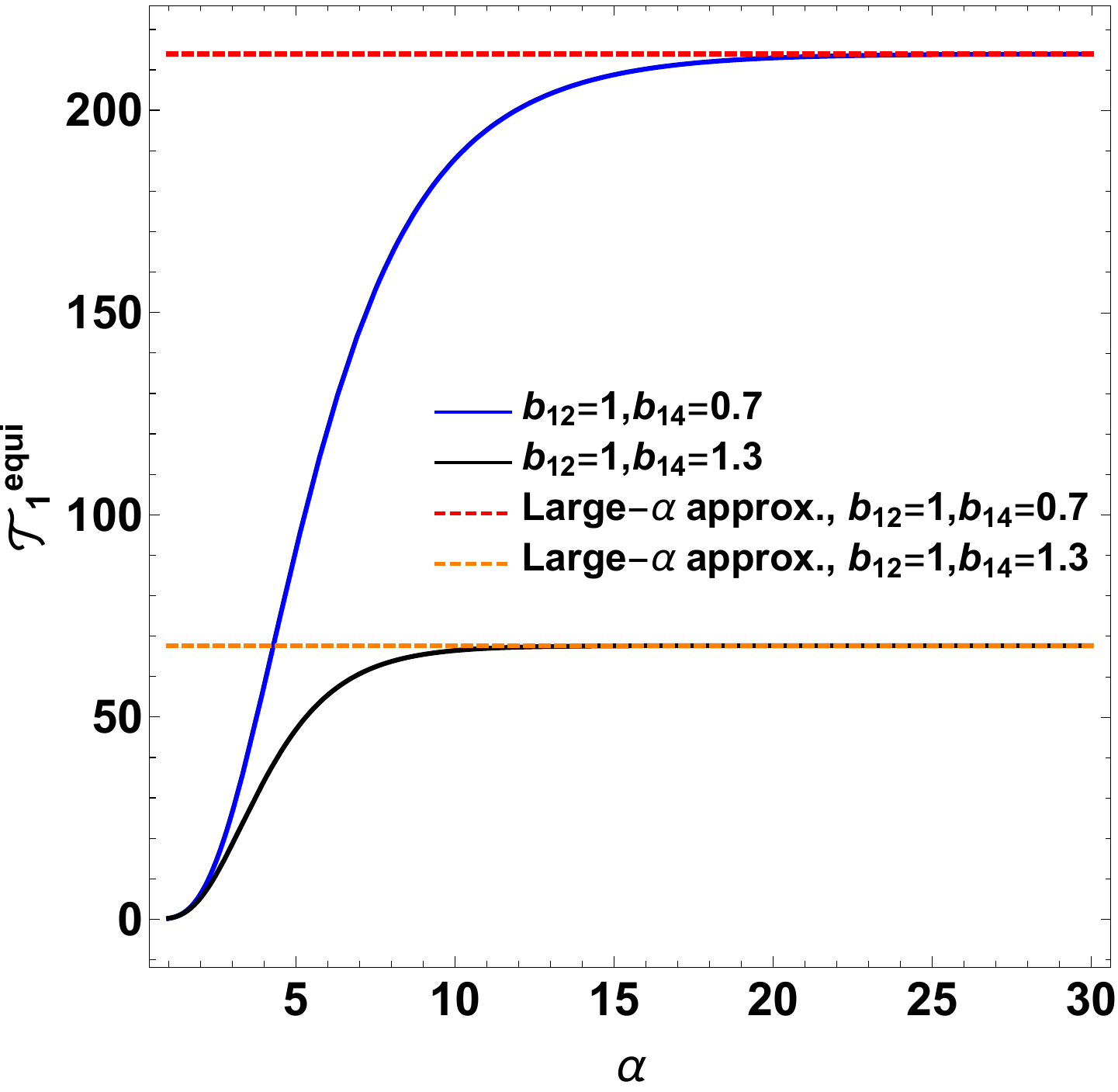}
	\includegraphics[width=0.48\linewidth]{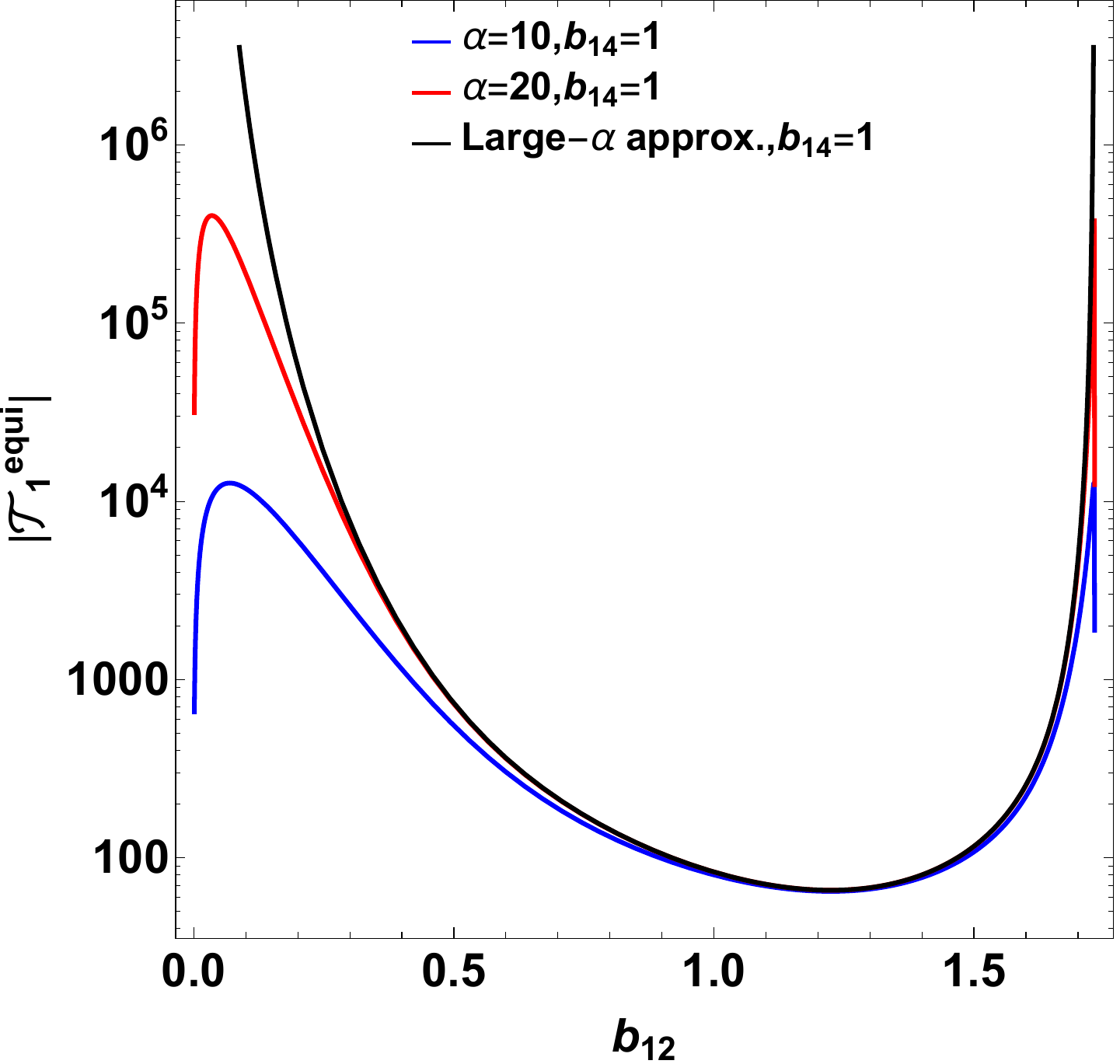}
	\caption{Equilateral trispectrum, ${\cal T}_{1}^\text{equi}$, as a function of: (left plot) $\alpha$ for 2 different values of $b_{14}$, (right plot) $b_{12}$ for two different values of $\alpha$. I also compare the full expressions with the large-$\alpha$ approximation. I fixed $ |c_s^2|=1, A=1$ in both plots.}
	\label{fig:T1equiang}
\end{figure}

For the equilateral shape, $k_1=k_2=k_3=k_4=k$, the result converges yielding
\begin{eqnarray} \label{K1 equi}
&& {\cal K}_1^\text{equi} \left(k, k,k, k\right)=  -2 \cdot 144 \left(\frac{A}{c_s^2}\right)^2 b_{12}  k^3 |c_s|^6  \times \nonumber \\ &&  \quad \quad  \left(\frac{4}{b_{12}^6}+\frac{4}{(b_{12}+2)^6}-\frac{8}{b_{12}^3 (b_{12}+2)^3}-b_{12} \frac{ 93 b_{12}^4-1048b_{12}^2+3600 }{64 \left(4-b_{12}^2\right)^3}\right) + \text{5 perm.} \, \, 
\end{eqnarray}
where $b_{ij}=k_{ij}/k_4$ ranges between zero and two. The result diverges precisely in the collinear and anti-collinear limits where the large-$\alpha$ approximation is no longer valid.
The remaining 5 permutations correspond to changes in the external legs which amount to swap $b_{12}$ by the remaining 5 possibilities. However, due to momentum conservation there are only 2 independent combinations ($b_{12}, b_{14}$) because $b_{12}=b_{34}, b_{13}=b_{24}, b_{14}=b_{23}$ and 
\begin{eqnarray}
k_{13} =\sqrt{k_1^2+k_2^2+k_3^2+k_4^2-k_{12}^2-k_{14}^2} \, .
\end{eqnarray}
In terms of $\cT$ the equilateral shape yields
\begin{eqnarray}
{\cal T}_1^\text{equi}\left(k, k,k, k\right)=  -  \left(\frac{1}{|c_{s}^{2}|}+1\right)^2  \frac{1}{ 2^9 k^3 |c_s|^2}  {\cal K}_1^\text{equi}\left(k, k,k, k\right)  \, .
\end{eqnarray}
In fig. \ref{fig:T1equiang} I plot the results as a function of $\alpha$ and $b_{12}$. As already mentioned, when $b_{12}$ or $b_{14}$ approach the (anti-) colinear limit the results approach those for the collapsed shapes where the trispectrum grows with $\alpha$.

\paragraph{Terms in ${\cal K}_2 $ and $\cK_3$:}

 Inserting the mode functions in eqs. \eqref{K2} and \eqref{K3} yields
\begin{eqnarray}
&& {\cal K}_2 (k_1,k_2,k_3,k_4) = 48 A |c_s|^2  \int_{\alpha}^{0} \frac{d z_1}{z_1}  \int_{\alpha}^{z_1} \frac{d z_2}{z_2} \left[  \left( \q_1 \cdot \q_2 \right) z_1^3 z_2  \frac{k_1 k_2 k_{12} k_3^2}{k_4^4}   e^{-(p_1 z_2 +p_2 z_1 )}   \times \right. \nonumber \\ && \left. \quad  \left(e^{\frac{2 k_3 z_1}{k_4}}+e^{2 z_1}-2\right) \left(\left(k_2 z_2+k_4\right) \left(e^{\frac{2 k_1 z_2}{k_4}} \left(k_1 z_2-k_4\right)+\left(e^{\frac{2 k_{12} z_1}{k_4}}-e^{\frac{2 k_{12} z_2}{k_4}}+2\right) \left(k_1 z_2+k_4\right)\right) \right. \right. \nonumber \\ &&\left.  \left. \quad - e^{\frac{2 k_2 z_2}{k_4}} \left(k_1 z_2+k_4\right) \left(k_4-k_2 z_2\right)\right) +\text{ 2 perm.} \right] + \text{ 5 perm.}  \, , \\
&&{\cal K}_3 (k_1,k_2,k_3,k_4)  = 48 A |c_s|^2 \int_{\alpha}^{0} \frac{d z_1}{z_1}  \int_{\alpha}^{z_1} \frac{d z_2}{z_2}  \left[  \left( \q_1 \cdot \q_2 \right) z_2^3 z_1 \frac{k_1 k_2  k_{12} k_3^2}{k_4^4}  e^{-(p_1 z_2 +p_2 z_2 )}  \times \right.  \nonumber \\ && \left. \quad \times  \left(e^{\frac{2 k_2 z_1}{k_4}} \left(k_1 z_1+k_4\right) \left(k_4-k_2 z_1\right)+\left(k_2 z_1+k_4\right) \left(e^{\frac{2 k_1 z_1}{k_4}} \left(k_4-k_1 z_1\right)-2 \left(k_1 z_1+k_4\right)\right)\right) \times \right. \nonumber \\ && \left.  \quad  \left(-e^{\frac{2 k_{12} z_1}{k_4}}+e^{\frac{2 k_3 z_2}{k_4}}+e^{\frac{2 k_{12} z_2}{k_4}}+e^{2 z_2}-2\right)  + \text{2 perm.} \right] + \text{5 perm.}
\end{eqnarray}
In the flattened configuration, $k_1=k_2=k_3=k_4/3$, I again find that only the terms where $\cR_{k_4}$ is in the innermost commutator grow as $\alpha^6$. Among the 9 remaining permutations there are 2 distinct scalar products in ${\cal K}_2$: $\q_1 \cdot (-\q_{12}) = -2 k^2 + \text{5 perm.}\,, \q_1 \cdot \q_2 =  k^2 +\text{2 perm.}$ In the case of $\cK_3$ there are 3 distinct cases: $\q_1 \cdot (-\q_{14})= 2 k^2, \q_1 \cdot \q_4 = - 3 k^2, \q_4 \cdot (-\q_{14})= -6k^2$, each with 3 permutations $k_1 \rightarrow \{k_2, k_3\}$.  After taking that into account one finds
\begin{eqnarray}
{\cal K}_2^\text{flat} \left(k, k,k, 3k\right)
&=&    -\frac{80}{81} \left( \frac{A}{ |c_s^2|} \right)  \alpha^6 |c_s|^{4}  k^3  \, , \\
{\cal K}_3^\text{flat} \left(k, k,k, 3k\right)
&=&   -\frac{176}{81} \left( \frac{A}{ |c_s^2|} \right)  \alpha^6 |c_s|^{4}  k^3 \, .
\end{eqnarray}
The large-$\alpha$ approximation is again accurate and yields the trispectrum
\begin{eqnarray} \label{Trispectrum mixed term}
{\cal T}_\text{cross}^\text{flat} \left(k, k,k, 3k\right)&=& -\frac{ 1}{2^9 k^3  3^{3/4} |c_s|^2}   \left(\frac{1}{|c_{s}^{2}|}+1\right)^2  \left( {\cal K}_2^\text{flat} \left(k, k,k, 3k\right) +  {\cal K}_3^\text{flat} \left(k, k,k, 3k\right)\right)  \nonumber \\ & \simeq &\frac{A \alpha^6  }{162 \cdot 3^{3/4}}  \left(\frac{1}{|c_{s}^{2}|}+1\right)^2 \, .
\end{eqnarray}

In the equilateral limit, $k_1=k_2=k_3=k_4=k$, among the 18 permutations 12 involve scalar products of $\q_i \cdot (-\q_{ij})$ while the other 6 involve terms of the form $\q_i \cdot \q_{j}$. Nevertheless, because of momentum conservation, there are only 3 different angles. Therefore, the terms in $\q_i \cdot  (-\q_{ij})$ repeat 4 times while the ones on $\q_i \cdot \q_{j}$ repeat twice. The final expressions in the large-$\alpha$ limit are not very illuminating so I present them in eqs. \eqref{K2 equi} and \eqref{K3 equi} of the appendix. In fig. \ref{fig:T34equiang} I plot ${\cal T}_{2}^\text{equi}+{\cal T}_{3}^\text{equi}$ and compare it with the large-$\alpha$ approximations. Similarly to $\cK_1^\text{equi}$, in the collinear limit, $b_{12} \rightarrow \{0,2\}$, the trispectrum grows with $\alpha$. 
\begin{figure}
	\centering
	\includegraphics[width=0.48\linewidth]{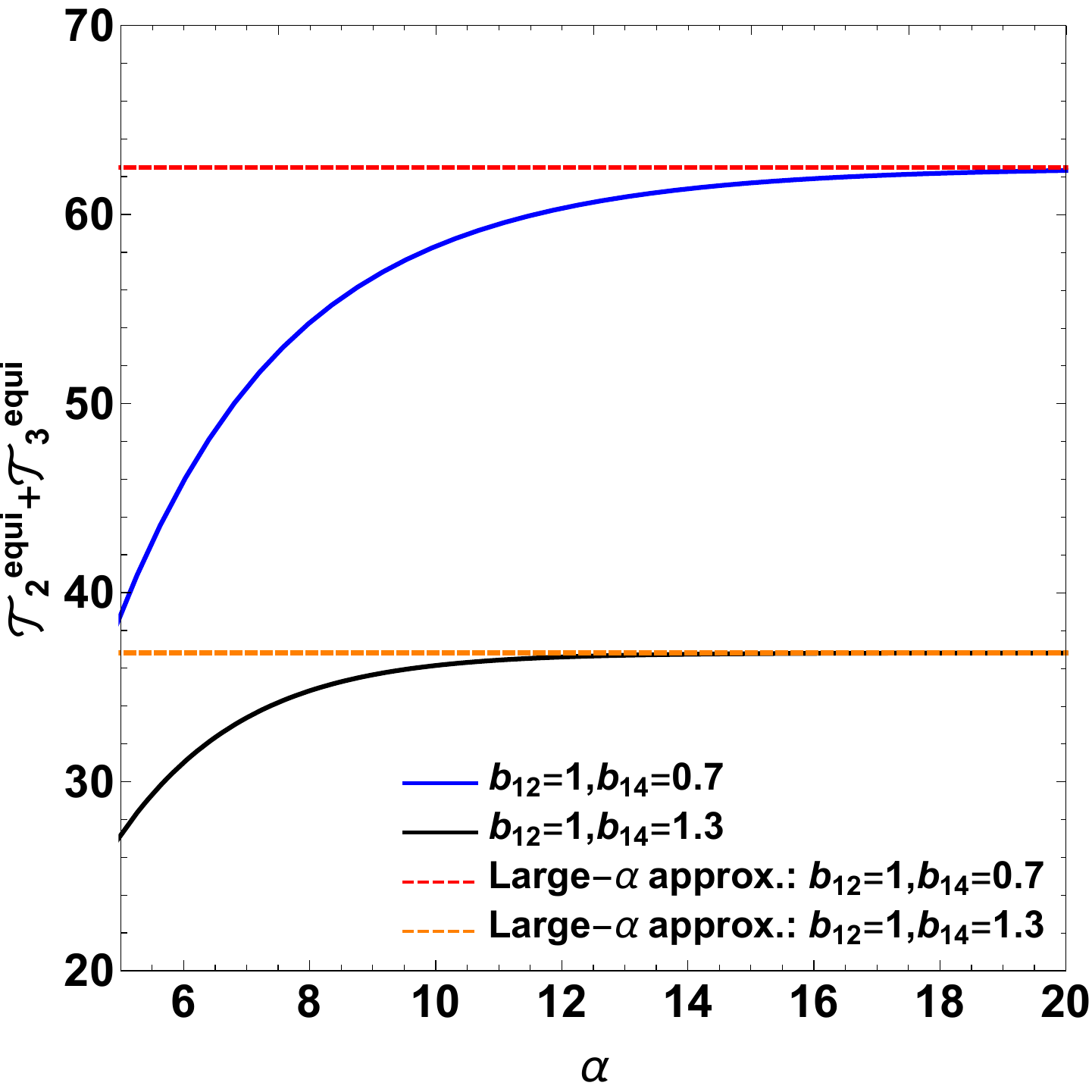}
	\includegraphics[width=0.48\linewidth]{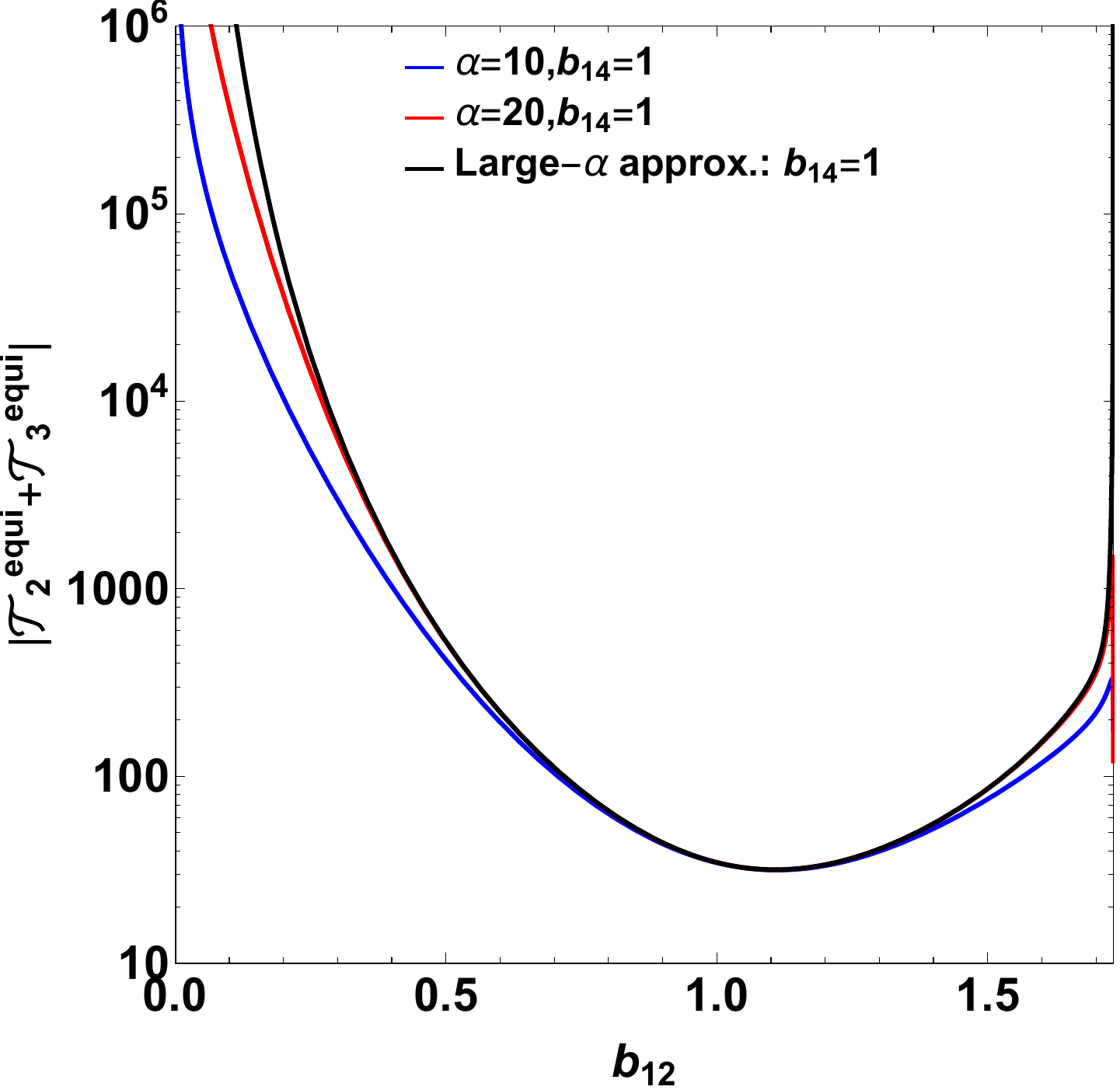}
	\caption{Equilateral trispectrum, ${\cal T}_{2}^\text{equi}+{\cal T}_{3}^\text{equi}$, as a function of: (left plot) $\alpha$ for 2 different values of $b_{14}$, (right plot) $b_{12}$ for two different values of $\alpha$. I also compare the full expressions with the large-$\alpha$ approximation. I fixed $ |c_s^2|=1, A=1$ in both plots.}
	\label{fig:T34equiang}
\end{figure}

\paragraph{Terms in ${\cal K}_4$:}

\begin{figure}
	\centering
	\includegraphics[width=0.48\linewidth]{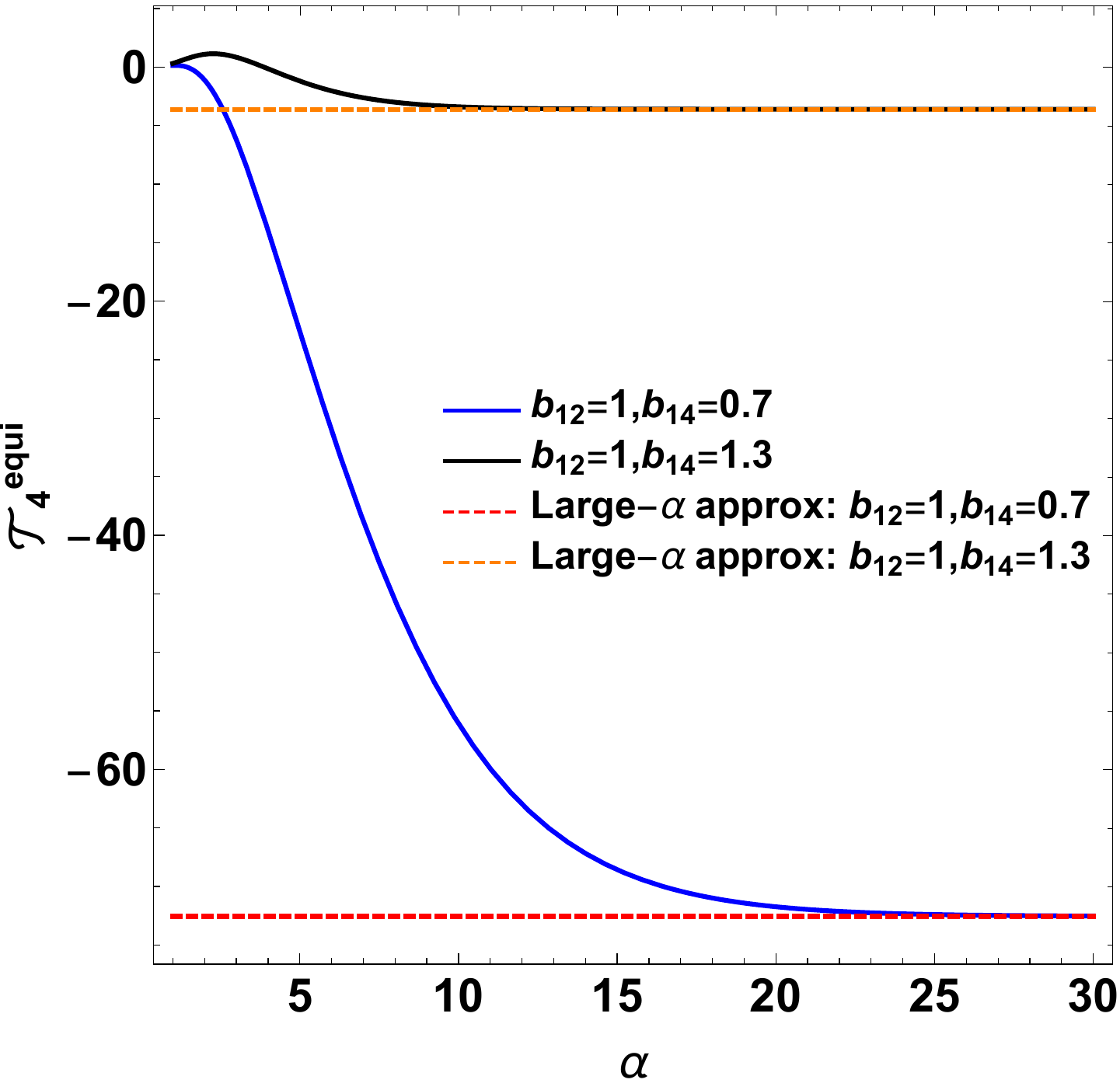}
	\includegraphics[width=0.48\linewidth]{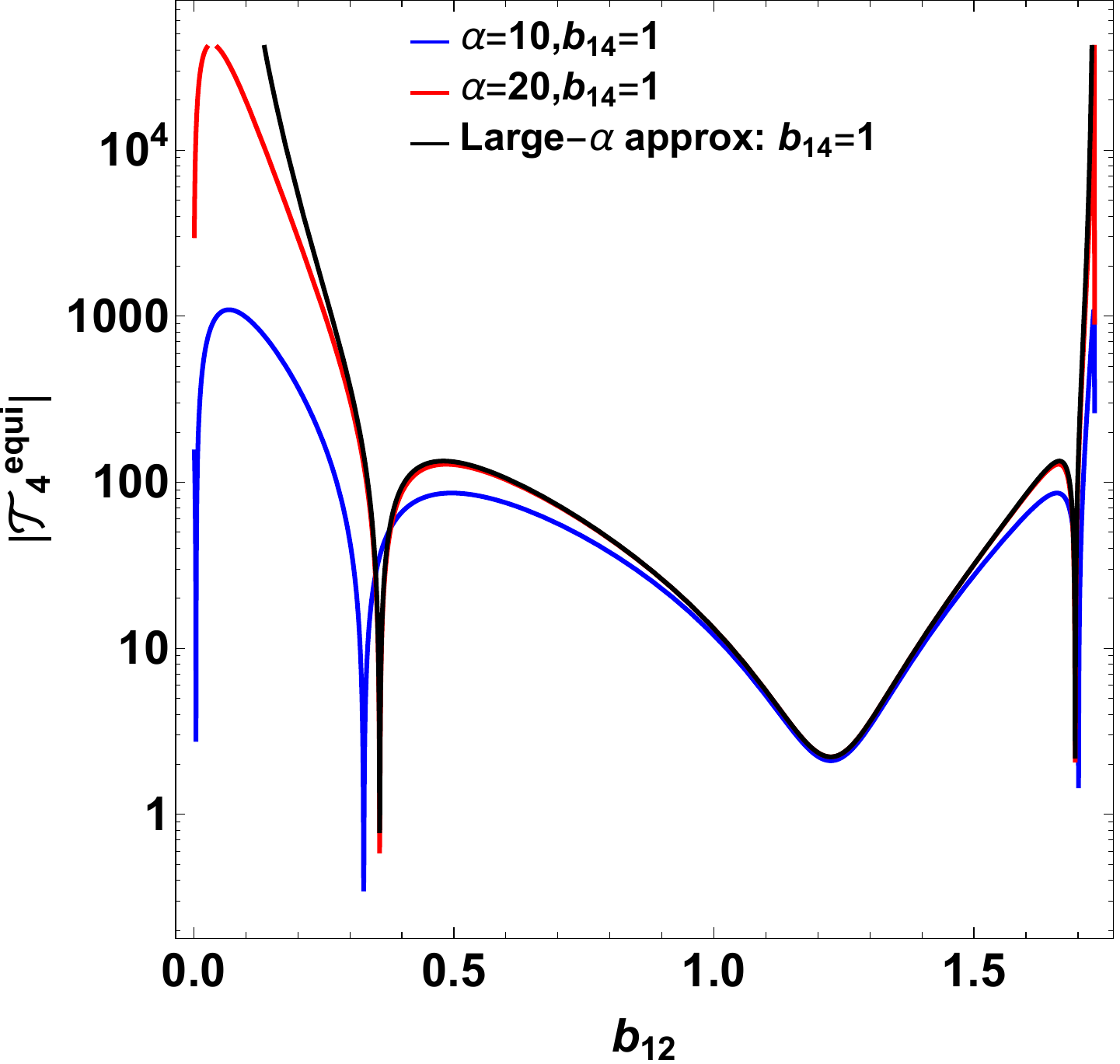}
		\caption{Equilateral trispectrum, ${\cal T}_{4}^\text{equi}$, as a function of: (left plot) $\alpha$ for 2 different values of $b_{14}$, (right plot) $b_{12}$ for two different values of $\alpha$. I also compare the full expressions with the large-$\alpha$ approximation. I fixed $ |c_s^2|=1, A=1$ in both plots.}
	\label{fig:T4equiang}
\end{figure}

Finally I evaluate the terms in ${\cal K}_4$. The expression for the integral in eq. \eqref{K4} after inserting the mode functions is long so I give it in eq. \eqref{K4 app} and only discuss here the flattened and equilateral shapes.
In the flattened case, $k_1=k_2=k_3=k_4/3$, among the leading 27 terms where $k_4$ is in the innermost commutator there are 6 different combinations:
\begin{eqnarray}
\begin{cases}
(\q_1 \cdot (-\q_{12}) ) (\q_3 \cdot \q_{12}) = -4 k^4\nonumber \\
(\q_1 \cdot (-\q_{12}) ) (\q_3 \cdot \q_4 ) = 6 k^4  \nonumber \\
(\q_1 \cdot (-\q_{12}) ) (\q_{12} \cdot \q_4 ) = 12 k^4  \nonumber 
\end{cases}   \text{+ 5 perm.} , \quad 
\begin{cases}
((\q_1 \cdot \q_2 ) (\q_3 \cdot \q_4 ) = -3 k^4 \nonumber \\
(\q_1 \cdot \q_2 ) (\q_3 \cdot \q_{12}) = 2 k^4 \nonumber \\
(\q_1 \cdot \q_2 ) (\q_{12} \cdot \q_4 ) =- 6 k^4 \nonumber 
\end{cases}  \text{+2 perm.} 
\end{eqnarray}
After summing the different contributions, the trispectrum in the flattened limit reads
\begin{eqnarray} \label{Trispectrum zeta' zeta2 }
\cT_4^\text{flat} (k,k,k,3k)=\frac{55}{ 7776 \cdot 3^{3/4}}  \alpha ^6\left(\frac{1}{|c_{s}^{2}|}+1\right)^2 \, .
\end{eqnarray}
In the equilateral case, $k_1=k_2=k_3=k_4$, there are 12 different combinations corresponding to:
\begin{eqnarray}
\begin{cases}
	2(\q_1\cdot \q_2 ) (\q_3 \cdot \q_4 ) = 2 k^4 \cos(\theta_{12})^2  \nonumber \\
4(\q_1\cdot (-\q_{12}) ) (\q_3 \cdot \q_4) =- 4 k^4 \cos(\theta_{12})(1+\cos(\theta_{12}))   \nonumber \\
4(\q_1\cdot \q_2 ) (\q_3 \cdot \q_{12} ) =  -4 k^4 \cos(\theta_{12})(1+\cos(\theta_{12}))   \nonumber \\
8(\q_1\cdot (-\q_{12}) ) (\q_3 \cdot \q_{12} ) =  8 k^4 (1+\cos(\theta_{12}))(1+\cos(\theta_{12})) 
\end{cases} \text{+ 2 perm.} 
\end{eqnarray}
I give the full expression for the trispectrum in the equilateral shape, $\cT_4^\text{equi}$, in eq. \eqref{K4 equi} and plot it in fig. \ref{fig:T4equiang}. The behavior is qualitatively similar to that of $\cT^\text{equi}_{1,2,3}$.

To summarize, all contributions of the exchange diagram to the trispectrum peak on certain collapsed shapes, the leading ones proportionally to $\alpha^{6}$, while other configurations give an $\alpha$-independent contribution in the large-$\alpha$ limit.

\section{Observational constraints \label{observational constraints}}

After having collected and derived a few predictions for the spectrum of scalar perturbations, I can now use the latest observational data to constraint the models described by the EFT with imaginary sound speed, i.e. models where the entropic directions have a large and negative effective mass.

The parameter $x$, appearing in the mode functions in eq. \eqref{mode function}, controls the amplitude and tilt of the power spectrum, while $\alpha$ controls the EFT cutoff and the size of the non-Gaussianities. Although these two parameters are related they might not be equal as it is clear from the case of rapid-turn attractors where $x= \alpha \pi(\sqrt{1+|c_s|^2}+1)/(2|c_s|)$ (see app. \ref{app: rapid-turn attractors}). Therefore, constraints on the power spectrum only restrict $x$ while non-Gaussian constraints only affect $\alpha$.

\subsection{Power spectrum and spectral index}

The amplitude and tilt of the scalar power spectrum at CMB scales are measured to be $P_\zeta^\text{obs}= 2.2 \times 10^{-9}, \, n_s^\text{obs} = 0.965$ \cite{Akrami:2018odb}. On small scales, the spectrum is further constrained by the overproduction of primordial black holes (PBH) \cite{Germani:2018jgr}
\begin{eqnarray}
P_\zeta\lesssim  P_\zeta^\text{PBH} \sim  10^{-3}-10^{-2} \, . \label{pbh constraint}
\end{eqnarray}
In the EFT described in sec. \ref{sec:Basics of the model} the scalar power spectrum for $x \gg 1$ is approximately given by eqs. \eqref{power spectrum} and \eqref{beta squared} \cite{Brown:2017osf,Garcia-Saenz:2018ifx}. If the $k$-dependence of $\rho$ and $\theta$ is much milder than that of $e^{2x}$, as it tends to be the case, then the spectral tilt is essentially controlled by $x$, namely,
\begin{eqnarray}
n_s-1 \simeq -2 \epsilon  - \delta + 2 \frac{\dot{x}}{H} \, . \label{spectral tilt obs} 
\end{eqnarray}
where $\delta \equiv \dot{\epsilon}/(\epsilon H)$. 
The PBH constraint is quite constraining if $x$ is proportional to some positive power of the first slow-roll parameter, $x = x_f \epsilon^n$. In that case, $x$ grows in time and exponentially enhances $P_\zeta$ on small scales. Using the fact that $\epsilon \simeq 1$ at the end of inflation, and assuming that $\rho, \theta$ do not vary much during inflation, eq. \eqref{pbh constraint} then implies that
\begin{eqnarray} \label{PBH constraint 2}
 P_\zeta^\text{obs} \, \epsilon_\text{CMB}  \, \exp\left[ 2(1-\epsilon_\text{CMB}^n) x_f\right]  \lesssim P_\zeta^\text{PBH}  \, .
\end{eqnarray}
Moreover, in such cases the spectral index is given, for $x \gg 1$, by
\begin{eqnarray}
n_s-1 \simeq -2 \epsilon + 2 n x \delta  \, . \label{spectral tilt 2}
\end{eqnarray} 
This requirement jointly with eq. \eqref{PBH constraint 2} strongly constraints the background dynamics. 
For example, in the case of hyperinflation \cite{Brown:2017osf} where $n=1/2$, eq. \eqref{PBH constraint 2} translates, using eqs. \eqref{x, cs vs omega, xi} and \eqref{epsilon}, into a constraint on the field-space curvature $L$
\begin{eqnarray} \label{PBH constraint 3}
\frac{M_p}{L} &\lesssim&
\frac{1}{ (2-\sqrt{2}) \sqrt{2} \pi (1-\sqrt{\epsilon_\text{CMB}})} \log \left( \frac{P_\zeta^\text{PBH}}{\epsilon_\text{CMB}  P_\zeta^\text{obs} }\right) \, .
\end{eqnarray}
More generally, if $x\propto \epsilon^n$ then from eqs. \eqref{PBH constraint 2} and \eqref{spectral tilt 2} models such that at CMB scales:
\begin{enumerate}
	\item   $\epsilon \lesssim - x \delta  $ (e.g. polynomial potentials) are excluded as they lead to a blue scalar spectrum.

\item $\epsilon \simeq 0.02 \gg 2nx\delta $ (e.g. exponential potentials) are generically outside the region of validity, $x \gg 1$, unless $n \ll 1$, because $x \lesssim 8.5 \times 0.02^n$.

\item $ \epsilon \ll 1, \, n x \delta \simeq {\cal O}(-0.01)$ (e.g. hilltop-like potentials) are still viable. In this case, by fine-tuning the initial conditions it is possible to satisfy all the constraints.

\end{enumerate}

\subsection{Non-Gaussianities}

In this section, I translate the observational constraints on the bispectrum and trispectrum into constraints on $\alpha$. The constraints apply to all models described by the EFT in sec. \ref{sec:Basics of the model}. How those constraints affect the parameters of the UV model will then vary in each case. In the end I particularize to the case rapid-turn attractors described in app. \ref{app: rapid-turn attractors} \cite{Brown:2017osf,Bjorkmo:2019fls}.
Note that the results for the bi and trispectrum in the collapsed limits should only be trusted at the order of magnitude level because of their  UV sensitivity, as discussed in the beginning of sec. \ref{sec:non-Gaussianities}.

\subsubsection*{Bispectrum}

I start by the bispectrum where a detailed study has been performed in \cite{Garcia-Saenz:2018vqf,Fumagalli:2019noh}. Here I review those results and make a direct comparison with the observational constraints to get a constraint $\alpha$. 

The bispectrum constraints are given in terms of $B_\Phi \left(k_{1}, k_{2}, k_{3}\right) = (3/5)^3 \, B \left(k_{1}, k_{2}, k_{3}\right) $ where $\Phi=3/5 \zeta$ is the Newtonian potential. Two characteristic templates are the equilateral and orthogonal respectively given by 
\begin{eqnarray}
&& B_\Phi^{\text {equil}}\left(k_{1}, k_{2}, k_{3}\right) = \\ &&  \quad 6 \Delta_\Phi^2  f_{\text {NL }}^{\text {equi}}  \left[-\frac{1}{(k_{1} k_{2})^{3}} -\frac{1}{(k_{2} k_{3})^{3}}-\frac{1}{(k_{3} k_{1})^{3}}-\frac{2}{\left(k_{1} k_{2} k_{3}\right)^{2 }} + \left[\frac{1}{k_{1} k_{2}^{2} k_{3}^{3}}+5 \text { perms. }\right] \right] \, , \quad \\
&& B_\Phi^{\text {orth}} \left(k_{1}, k_{2}, k_{3}\right)= \\ && \quad 6  \Delta_\Phi^2 f_{\mathrm{NL}}^{\text {orth}} \left[-\frac{3}{(k_{1} k_{2})^{3}}-\frac{3}{(k_{2} k_{3})^{3}}-\frac{3}{(k_{3} k_{1})^{3}}-\frac{8}{\left(k_{1} k_{2} k_{3}\right)^{2}}+  \left[\frac{3}{k_{1}k_{2}^{2} k_{3}^{3}}+5 \text{ perms. }\right]\right] 
\end{eqnarray}
where $\Delta_\Phi=\left( 3 \sqrt{2}\pi /5 \right)^2 P_{\zeta} $. When confronting the predictions with the observational data, optimally, one should construct a template dedicated  to the scenario under scrutinity. However, as it was shown in \cite{Meerburg:2019qqi,Garcia-Saenz:2018vqf} the orthogonal template is strongly correlated with the bispectrum generated in this model for large $\alpha$. Therefore, I can use the observational constraint $f_\text{NL}^\text{orth} = -38 \pm 24$ (68 \% CL) \cite{Akrami:2019izv} to place constraints on $\alpha$. More concretely, using eqs. \eqref{shape function}, \eqref{shape function zeta'3} and \eqref{shape function zeta'zeta^2} and the fact that in the flattened limit $ B^{\text {orth}}_\Phi (2k,k,k)= - 3 \Delta_\Phi^2  f_{\mathrm{NL}}^{\text {ortho}}/ k^6 $ one finds that
\begin{equation} \label{fNL orthogonal}
f_{N L}^{\text {ortho }} \simeq - \frac{5}{288} (A+1)\left(\frac{1}{|c_s^2|}+1\right) \alpha^3 \, .
\end{equation}
The parameters $A$ and $c_s$ depend on the microphysics of each model. However, unless $A$ is close to $-1$, the absolute value of $f_\text{NL}$ is roughly bounded from below and essentially depends on $\alpha$. Thus, imposing the $2\sigma$ constraint $-86<f_{N L}^{\text {ortho }}<10$ \cite{Akrami:2019izv} one finds
\begin{eqnarray} \label{bispectrum constraint on alpha}
-8\lesssim\alpha (A+1)^{1/3} \left(\frac{1}{|c_s^2|}+1\right)^{1/3} \lesssim 17  \, .
\end{eqnarray}
For rapid-turn models, where $\alpha^2= 4 |c_s|^2/(1+|c_s|^2)\omega^2$ and $\omega$ is the turning rate in field space defined in eq. \eqref{w}, it implies that
\begin{eqnarray} \label{bispectrum constraint on omega}
-4 \lesssim \omega |c_s|  (A+1)^{1/3} \left(\frac{1}{|c_s^2|}+1\right)^{-1/6} \lesssim 9\, ,
\end{eqnarray}
which, in the particular case of hyperinflation ($A\simeq -1/3$, $|c_s^2|=1$) \cite{Fumagalli:2019noh,Garcia-Saenz:2019njm}, imposes $\omega \lesssim 11$.

\subsubsection*{Trispectrum} 

In sec. \ref{sec:trispectrum} I derived the contributions to the trispectrum from the contact interaction and the exchange diagram in fig. \ref{fig:4ptX}. I derived the general expressions and then focused on two particular shapes: the equilateral, $k_1=k_2=k_3=k_4=k$, and the flattened, $k_1=k_2=k_3=k_4/3=k$. In this section, I focus on the flattened as it the representative of the leading contributions, and derive consequent constraints on $\alpha$. 

From the contact term eqs. \eqref{T zeta'4}, \eqref{T zeta'2 dzeta2} and \eqref{T dzeta4} give the trispectrum
\begin{eqnarray} \label{contact trispectrum collapsed shape}
\cT^\text{flat}_\text{contact} =\frac{\alpha^5}{ 180 \cdot 3^{3/4}}  \left(\frac{1}{|c_{s}^{2}|}+1\right)  \left(B  |c_s|^2    - C  +  \frac{1}{4}  \right)  \, ,
\end{eqnarray}
while from the exchange diagram eqs. \eqref{Trispectrum zeta'3}, \eqref{Trispectrum mixed term} and \eqref{Trispectrum zeta' zeta2 } yield
\begin{eqnarray} \label{exchange trispectrum collapsed shape}
\cT^\text{flat}_\text{exchange}=\frac{\alpha^6}{162\cdot 3^{3/4}}  \left(\frac{1}{|c_{s}^{2}|}+1\right)  \left( \frac{3}{16 }  A^2   + A   + \frac{55}{ 48}    \right) \, .
\end{eqnarray}
The current observational constraints are given in terms of parameters associated with local and contact interactions. At $68\%$ CL they read\footnote{For $g_\text{NL}^{\dot{\sigma}^4} , g_\text{NL}^{(\partial \sigma)^4} $ I used the constraint in the first line of table 24 of \cite{Akrami:2019izv}.} \cite{Akrami:2019izv}
\begin{eqnarray} \label{constraint on gNL}
&g_\text{NL}^\text{local}  = (-5.8 \pm 6.5) \times 10^4 \, , \quad  g_\text{NL}^{\dot{\sigma}^4}  =  (-0.8 \pm 1.9) \times 10^6 \, , &  \\  &g_\text{NL}^{(\partial \sigma)^4} = (-3.9 \pm 3.9) \times 10^5 \, .
\end{eqnarray}
Contributions to local non-Gaussianities are typically small in these models due to the decay of the isocurvature mode on superhorizon scales. Therefore, I disregard $g_\text{NL}^\text{local}  $ and focus instead on $g_\text{NL}^{\dot{\sigma}^4}$ and $g_\text{NL}^{(\partial \sigma)^4}$ which are defined through \cite{Senatore:2009gt}
\begin{eqnarray} \label{gNL}
\left\langle\zeta_{k_{1}} \zeta_{k_{2}} \zeta_{k_{3}} \zeta_{k_{2}}\right\rangle' &=& (2\pi)^3 \frac{221184}{25} g_\text{NL}^{\dot{\sigma}^4} P_\zeta^3 \frac{1}{k_{1} k_{2} k_{3} k_{4} K^{5}}  \\
\left\langle\zeta_{k_{1}} \zeta_{k_{2}} \zeta_{k_{3}} \zeta_{k_{2}}\right\rangle' &=& (2\pi)^3 \frac{165888}{2575} g_\text{NL}^{(\partial \sigma)^4} P_{\zeta}^{3}  \left(\frac{2 K^{4}-2 K^{2} \Sigma k_{i}^{2}+K \sum k_{i}^{3}+12 k_{1} k_{2} k_{3} k_{4}}{k_{1}^{3} k_{2}^{3} k_{3}^{3} k_{4}^{3} K^{5}} \right) \times 
 \nonumber \\ && \left[\left(\q_{1} \cdot \q_{2}\right)\left(\q_{3} \cdot \q_{4}\right)+\text {2 perm.}\right] \, , 
\end{eqnarray}
where $K=\sum_i k_i$ and the prime in the correlator denotes that the delta function was suppressed\footnote{Note that I use a different Fourier convention and so the correlator differs by a factor of $(2\pi)^3$ compared to the notation in \cite{Akrami:2019izv}.}.
Unfortunately, these templates are not optimized for contributions from exchanged diagrams which tend to be subdominant, contrarily to what happens in the models studied here.
In light of that, I will first constraint $\alpha$ using the contribution to the trispectrum from the contact interaction in eq. \eqref{contact trispectrum collapsed shape}. This contribution should have a larger correlation with the templates associated with $ g_\text{NL}^{\dot{\sigma}^4}, g_\text{NL}^{(\partial \sigma)^4}$. Afterwards, I use the same constraint on $g_\text{NL}^{\dot{\sigma}^4}$ as a proxy to constraint the exchange diagram.
Namely, by imposing the 2$\sigma$ constraint on $g_\text{NL}^{\dot{\sigma}^4}$ in eq. \eqref{constraint on gNL} as a benchmark\footnote{Using instead the constraint on $g_\text{NL}^{(\partial \sigma)^4}$ gives similar results.} I find, using eq. \eqref{contact trispectrum collapsed shape}, that
\begin{eqnarray} \label{trispectrum constraint from contact interaction}
\frac{675}{275 \cdot 3^{9/4}} \left| \cT^\text{near-flat}\right|   \lesssim  5\times 10^6  \quad  \Rightarrow \quad  \alpha \lesssim  98\left(B  |c_s|^2    - C  +  \frac{1}{4 }\right)^{-1/5}   \, .
\end{eqnarray}
Repeating the same steps for the contribution from the exchange diagram in eq. \eqref{exchange trispectrum collapsed shape}, bearing in mind the caveats mentioned above, yields
\begin{eqnarray} \label{trispectrum constraint from exchange diagram}
 \alpha & \lesssim  & 34.4   \left[ \left(\frac{1}{|c_{s}^{2}|}+1\right)  \left(\frac{3}{16 }  A^2   + A   + \frac{55}{ 48}\right) \right]^{-1/6} \, .
\end{eqnarray}
By comparing eqs. \eqref{trispectrum constraint from contact interaction}, \eqref{trispectrum constraint from exchange diagram} with eq. \eqref{bispectrum constraint on alpha},
I conclude that for $A,B,C,c_s \sim {\cal O}(1)$ the bispectrum currently provides the strongest constraints.

\section{Conclusion}

In this work, I studied inflationary models described by a single-field EFT with an imaginary speed of sound \cite{Garcia-Saenz:2018ifx}.
Such EFT arises when the effective entropic mass(es), $m_s$, are large and negative. Models where the field-space metric is hyperbolic and a large turning rate in field-space counterbalances the instability are among the interesting examples  \cite{Renaux-Petel:2015mga,Brown:2017osf,Mizuno:2017idt,Christodoulidis:2018qdw,Garcia-Saenz:2018ifx,Bravo:2019xdo,Aragam:2019khr,Garcia-Saenz:2019njm,Bjorkmo:2019fls, Chakraborty:2019dfh,Christodoulidis:2019jsx,Bjorkmo:2019aev}. Slow-roll inflation is then still possible, even in steeper potentials, albeit in a different attractor. 
The spectrum of scalar perturbations also has interesting new features. Its amplitude is enhanced by $\sim e^{2x}$ where $x$ is closely related to $m_s/H$ \cite{Garcia-Saenz:2018ifx,Garcia-Saenz:2018vqf}. 
Interestingly, non-Gaussian parameters only have a polynomial dependence on $m_s/H$ \cite{Garcia-Saenz:2018vqf,Fumagalli:2019noh,Bjorkmo:2019qno}.

In sec. \ref{sec:non-Gaussianities} I extended the discussion of non-Gaussianities in these models. First, I reviewed the bispectrum calculation in sec. \ref{sec:bispectrum}. Then, in sec. \ref{sec:trispectrum} I completed the trispectrum calculation by computing contributions from both the contact interaction and the exchange diagram in fig. \ref{fig:4ptX}. So far only one term had been computed \cite{Fumagalli:2019noh}. I found that for most shapes the bi and trispectrum are constant in the limit $\alpha= |c_s m_s|/H \gg 1$. However, for some configurations where all momenta collapse to a line (e.g. the flattened shapes in fig. \ref{fig:flat3pf}) the time integrals become dominated by the UV cutoff and proportional to $\alpha^3$ in the bispectrum  (eqs. \eqref{shape function zeta'3}, \eqref{shape function zeta'zeta^2}). In the trispectrum, the dominant contribution, proportional to $\alpha^6$ (eq. \eqref{exchange trispectrum collapsed shape}), comes from the exchange diagram while the contact interaction gives a contribution proportional to $\alpha^5$ (eq. \eqref{contact trispectrum collapsed shape}), as anticipated in \cite{Fumagalli:2019noh}. When overlapping, the results found in this work agree quantitatively and qualitatively with the literature. Only the contribution to the trispectrum in eq. \eqref{Trispectrum zeta'3} differs a factor of 3 from the result found in \cite{Fumagalli:2019noh}. The UV sensitivity for the flattened shapes is typical of theories with excited initial states \cite{Chen:2006nt,Chen:2009bc} and implies that a precise calculation of non-Gaussianities is only achievable in the full multi-field system. Nevertheless, as I argued in sec. \ref{sec:non-Gaussianities}, if the fields quickly approach the massless and weakly coupled limit for scales above the cutoff, which is typically the case, they will be described by plane waves whose rapid oscillations exponentially suppress the contribution above the cutoff. Therefore, the results for the flattened shapes within the EFT still provide good order of magnitude estimates. 

Finally, in sec. \ref{observational constraints} I confronted the different predictions for the spectrum of scalar perturbations against observations. I pointed out that if $x$ is proportional to some positive power of the slow-roll parameter $\epsilon$, as in the case of the rapid-turn attractors like hyperinflation \cite{Brown:2017osf} or sidetracked inflation \cite{Garcia-Saenz:2018ifx}, the spectrum grows exponentially on small scales and so might overproduce primordial black holes. This provides a strong constraint for scenarios where the spectral index is controlled by $\epsilon$.  For example, for hyperinflation, the combination of a red spectral tilt at CMB scales with the PBH constraint excludes exponential and polynomial-like potentials.

I then used the observational constraints on non-Gaussianities \cite{Akrami:2019izv} to constrain the EFT parameter $\alpha$. For the bispectrum, I used the constraint on $f_\text{NL}^\text{ortho}$ whose template has been shown to strongly correlate with the bispectrum in this model \cite{Meerburg:2009fi,Garcia-Saenz:2018vqf}. I found in eq. \eqref{bispectrum constraint on alpha} the constraint on $\alpha$ and in eq. \eqref{bispectrum constraint on omega} I translated it in terms of the parameter $\omega$ which controls the turning-rate in field space in rapid-turn attractors. For hyperinflation, I found that $\omega \lesssim 11$ which strongly constraints the model.
Regarding the trispectrum, I used the constraints on the parameters $g_\text{NL}^{ \dot{\sigma}^4}$, $g_\text{NL}^{(\partial \sigma)^4}$ \cite{Akrami:2019izv} to constraint the contribution generated from the contact interaction, eq. \eqref{trispectrum constraint from contact interaction}, and from the exchange diagram,  eq. \eqref{trispectrum constraint from exchange diagram}. For the exchange diagram, which is the dominant contribution for large $\alpha$, a dedicated analysis is required as none of the associated templates is expected to be a good fit. Still, I used the same constraints as a proxy. In both cases, I found that the constraints on $\alpha$ from the trispectrum are weaker than that coming from the bispectrum. However, a dedicated analysis is required to make more definite statements.

To conclude, I  derived different constraints on the models described by the EFT with imaginary sound speed. Generically, I found that although constrained these models are still observationally viable if the parameter controlling the effective mass is roughly in the window $\alpha \lesssim 10-20$. Furthermore, if the parameter $x$ is proportional to a positive power of $\epsilon$ the only viable models are likely those where the spectral tilt is controlled by $2\dot{x}/H \simeq {\cal O}(-0.01)$,  like in hill-top potentials.

\section*{Acknowledgments}
I would like to thank David M.C. Marsh and Theodor Bjorkmo for several discussions regarding hyperinflation and rapid-turn attractors and, in particular, David M.C. Marsh for discussions related to the trispectrum computation. I would also like to thank Sebastien Renaux-Petel and Jacopo Fumagali for comments on a draft of this paper.

\appendix

\section{Rapid-turn attractors \label{app: rapid-turn attractors}}

A negatively curved field-space metric can potentially destabilize the inflationary trajectory  \cite{Renaux-Petel:2015mga,Brown:2017osf,Mizuno:2017idt,Christodoulidis:2018qdw,Garcia-Saenz:2018ifx,Bjorkmo:2019aev,Bravo:2019xdo,Aragam:2019khr,Garcia-Saenz:2019njm,Bjorkmo:2019fls, Chakraborty:2019dfh,Christodoulidis:2019jsx}. This instability can, however, be counterbalanced by a large angular velocity in field-space thus allowing for inflation in a new attractor solution. Recently, it has been argued that all such models follow the same rapid-turn attractor where the turning rate in field-space is characterized by \cite{Bjorkmo:2019fls}
\begin{eqnarray} \label{w}
\omega = || {\cal D}_t (\dot{\phi}^a/\dot{\phi})||/H \, .
\end{eqnarray} 
The superscript $a$ runs over the $N$-dimensional field-space, $\dot{\phi}=|| \dot{\phi}^a||$ and ${\cal D}_t$ is a covariant time derivative in the field-space metric. The large turning rate effectively generates a large and negative mass, $m_s^2= (\xi -1) \omega^2 H^2$, for the entropic direction. This direction can then be integrated out leading to the single-field EFT described in sec. \ref{sec:Basics of the model} with \cite{Bjorkmo:2019qno}
\begin{eqnarray} \label{x, cs vs omega, xi}
 x \simeq (2- \sqrt{3+\xi} ) \pi \omega/2\, ,\quad  |c_s|=\sqrt{(1-\xi)/(3+\xi)}\, ,
\end{eqnarray} 
where $\xi <1$. Below I briefly present two examples of such attractors with hyperbolic geometries: hyperinflation \cite{Brown:2017osf} and sidetracked inflation \cite{Garcia-Saenz:2018ifx}.

\begin{itemize}
	\item 
{\bf Hyperinflation:}  
It is described by the Lagrangian \cite{Brown:2017osf} 
\begin{eqnarray}
{\cal L} = - \frac 1 2 \left(\partial \phi\right)^2 - \frac 1 2 \sinh^2(\phi/L) \left(\partial \psi \right)^2 - V(\phi) \, ,
\end{eqnarray}
The new attractor solution is characterized by 
\begin{eqnarray}
\dot{\phi}=-3 H L \, , \quad   \dot{\psi} \sinh \frac{\phi}{L} =\sqrt{L \partial_{\phi} V-(3 H L)^{2}} \equiv \omega L H \, 
\end{eqnarray} 
with slow parameters given in the large $\omega$ limit by
\begin{eqnarray} \label{epsilon}
\epsilon = - \frac{\dot{H}}{H^2}  \simeq \frac 1 2 \omega^2 \left(\frac{L}{M_p}\right)^2 \, ,   \qquad  \eta =\frac{L V_{, \phi \phi}}{V_{, \phi}}  \, .
\end{eqnarray}
with $\omega L \ll M_p$. It is then clear that inflation can unfold in steeper potentials. The effective entropic mass is in this case $m_s^2=-2 \omega^2 H^2$. Therefore, for $\omega \gg 1$ the perturbations can be described by the EFT of sec. \ref{sec:Basics of the model}. The power spectrum is given by eq. \eqref{spectral tilt obs} with $2\rho \sin \theta =1$ while the spectral tilt is given by \cite{Mizuno:2017idt}
\begin{eqnarray}
n_s-1= -2 \epsilon + x \eta \, .
\end{eqnarray}

\item {\bf Sidetracked inflation:}
The system is characterized by the Lagrangian \cite{Garcia-Saenz:2018ifx}
\begin{eqnarray}
{\cal L} = - \frac 1 2 \left(1+ \frac{2 \psi^2}{M^2} \right) \left(\partial \phi\right)^2 - \sqrt{2} \frac{\psi}{M}\partial \phi \partial \psi - \left(\partial \psi \right)^2 - V(\phi, \psi) \, ,
\end{eqnarray}
where $V(\phi,\psi)=V_1(\phi)+m^2\psi^2/2$ and $m \gg H$. In the attractor solution $\psi \sim H$ and the entropic mass is
\begin{equation}
\frac{m_{s}^{2}}{H^{2}} \simeq 12 \frac{m}{H} \frac{\psi^{2}}{M^{2}} \operatorname{sign}\left(V_{, \varphi}\right) \,. 
\end{equation}
Therefore, if $V'<0$ then $ |m_s|/H \gg 1$ and the theory can be described by the EFT with imaginary sound speed given by $c_s^2 \simeq 3H/(2 m) \operatorname{sign}\left(V_{, \varphi}\right)$.
\end{itemize}

\section{Interaction Hamiltonians \label{app:interaction Hamiltonians }}

As we discussed in the previous sections the system under discussion can be well described by a single field EFT. Therefore, to leading order in the slow-roll parameters the system will inherit the interactions derived in the context of the EFT of single-field inflation \cite{Cheung:2007st,Garcia-Saenz:2018vqf}. 

\subsection{Cubic order}
The interaction Hamiltonian of cubic order to leading order in slow-roll and derivatives is given by \cite{Cheung:2007st} 
\begin{eqnarray} \label{cubic Hamiltonian}
H_{I,3} (\tau) = -\int d^{3} x \frac{a \epsilon M_{\mathrm{Pl}}^{2}}{H}\left(\frac{1}{c_{s}^{2}}-1\right) \left[\zeta^{\prime}(\vec{\nabla} \zeta)^{2}+\frac{A}{c_{s}^{2}} \zeta^{\prime 3}\right] \,.
\end{eqnarray}
For two-field models within the validity of the EFT, $A$ is given by \cite{Garcia-Saenz:2019njm} 
\begin{eqnarray}
A&=&-\frac{1}{2}\left(1+c_{s}^{2}\right)+\frac{2}{3}\left(1+2 c_{s}^{2}\right) \frac{\epsilon H^{2} M_p^2 R_\text{fs}}{m_{s}^{2}}-\frac{1}{6}\left(1-c_{s}^{2}\right)\left(\frac{\kappa V_{; s s s}}{m_{s}^{2}}+\frac{\kappa \epsilon H^{2} M_{\mathrm{Pl}}^{2} R_{\mathrm{fs}, s}}{m_{s}^{2}}\right) \, .
\end{eqnarray}
where $\kappa=\sqrt{2\epsilon}M_p/\eta_\perp$. For hyperinflation $A \simeq -1/3$ \cite{Garcia-Saenz:2019njm}.
After Fourier transforming the fields according to 
\begin{eqnarray} \label{Fourier convention}
\zeta(x) =  \int  \frac{d^3k}{(2\pi)^{3/2}} e^{i k x} \left[ a_k \zeta_k +a^\dagger_{-k} \zeta_k^{*} \right] , \, \quad 
\left[a_k,a_q^\dagger\right] = \delta^{(3)}(k-q) \,  ,
\end{eqnarray}
where $a,a^\dagger$ are the creation and annihilation operators and using $a\simeq -1/(\tau H)$, $P_{\zeta,0}= H^2/(8\pi^2 |c_s| \epsilon M_p^2)$ and eq. \eqref{cR definition}, the interaction Hamiltonian becomes of the form
\begin{eqnarray} \label{H3 decomposition}
H_{I,3} (\tau)  &=& h \left(\tilde{H}_{3,1}+\tilde{H}_{3,2}\right) \, ,
\end{eqnarray}
where
\begin{eqnarray}
h&\equiv &  \left(\frac{1}{|c_{s}^{2}|}+1\right)\frac{1}{8|c_s| } \sqrt{\frac{P_{\zeta,0}}{8\pi}}  \, , \label{h}\\
\tilde{H}_{3,1} &=& \frac{ \left(\q_1 \cdot \q_2\right) }{\tau} \int \left(\prod_{i=1}^{3} d^{3} k_i \right)  \delta^{(3)}(k_1+k_2+k_3) \mathcal{R}_{k_1}   \mathcal{R}_{k_2} \mathcal{R}'_{k_3} \, ,\\  \tilde{H}_{3,2}  &=& \frac{A}{\tau |c_s^2|} \int \left(\prod_{i=1}^{3} d^{3} k_i \right)  \delta^{(3)}(k_1+k_2+k_3)  \mathcal{R}^{\prime}_{k_1} \mathcal{R}^{\prime}_{k_2}\mathcal{R}^{\prime}_{k_3} \, ,
\end{eqnarray}
and I used the fact that $c_s^2<0$.

\subsection{Quartic order}

Up to 4-derivatives and to lowest order in the slow-roll parameters the interaction Hamiltonian to fourth order has the form \cite{Chen:2009bc}\footnote{Note that there is a factor of $a$ different from \cite{Chen:2009bc} because here the Hamiltonian is in conformal time.}
\begin{eqnarray}
 H_{I,4} (\tau) =  \int  \frac{d^{3} x}{H^{4}} \left[ \left(-\mu+9 \frac{\lambda^{2}}{\Sigma}\right) \zeta'^{4}+\left(3 \lambda c_{s}^{2}-\Sigma\left(1-c_{s}^{2}\right)\right)\left(\partial \zeta_{I}\right)^{2} \zeta'^{2} -\frac{c_s^2}{4} \Sigma\left(c_{s}^{2} -1\right)\left(\partial \zeta \right)^{4} \right] \, , \nonumber
\end{eqnarray}
where $\Sigma = M_p^2 H^2 \epsilon/c_s^2$ \cite{Seery:2005wm}. The terms in $\lambda$ and $\mu$ need to be derived from the full multi-field system as they cannot be determined uniquely from background quantities and the speed of sound. I leave the coefficients of those terms $B,C$ undetermined and define the Hamiltonian as
\begin{equation} \label{quartic Hamiltonian}
\begin{aligned} H_{I,4} (\tau) &= \int d^{3} x \left(\frac{1}{c_{s}^{2}}-1\right) \frac{\epsilon M_p^2}{ H^{2}} \left[ B \, \zeta'^{4}+C \, \left(\partial \zeta \right)^{2} \zeta'^{2} - \frac{c_s^2}{4} \left(\partial \zeta \right)^{4} \right] \, .\end{aligned}
\end{equation}
In Fourier space, and in terms of $\mathcal{R}$ defined in eq. \eqref{Dimensionless variable}, the Hamiltonian becomes
\begin{eqnarray}
 && H_{I,4} (\tau) =  \left(\frac{1}{c_{s}^{2}}-1\right)  \frac{  P_{\zeta,0}}{32 (2\pi) |c_s| } \int \left(\prod_{i=1}^{4} d^{3} k_i\right) \delta^{(3)} \left( \sum_{i=1}^4 k_i \right)  \times \nonumber \\ && \times  \left[ B \, \cR^{\prime}_{k_1} \cR^{\prime}_{k_2}\cR^{\prime}_{k_3} \cR^{\prime}_{k_4} -C \, \left(\q_1 \cdot \q_2 \right) \cR_{k_1} \cR_{k_2} \cR^{\prime}_{k_3} \cR^{\prime}_{k_4}- \frac{c_s^2}{4}\left( \q_1 \cdot \q_2 \right) \left(\q_3 \cdot \q_4 \right)\cR_{k_1} \cR_{k_2} \cR_{k_3} \cR_{k_4} \right]  , \nonumber \\  && \qquad \quad \, \,  \equiv \left(\frac{1}{c_{s}^{2}}-1\right)  \frac{  P_{\zeta,0}}{32 (2\pi) |c_s| } \int \left(\prod_{i=1}^{4} d^{3} k_i\right) \delta^{(3)} \left( \sum_{i=1}^4 k_i \right)  \times \sum_{i=1}^3 \tilde{H}_{4,i}  \label{tilde H4} \, .
\end{eqnarray}

\section{Trispectrum: general formulas \label{app:equilateral trispectrum}}
In this section, I gather different results obtained in the trispectrum computation which although useful their long form is not very illuminating. 
\subsection{Contact interaction: contributions from ${\cal I}_3$}
The first expression is the result of the integral in eq. \eqref{I3 int}
\begin{eqnarray} \label{I3 app}
&& \mathcal{I}_3(k_1,k_2,k_3,k_4)= \frac{|c_s|}{k_1^4} \left( \q_1 \cdot \q_2 \right) \left(\q_3 \cdot \q_4 \right)  \times  \\ && \quad  \left[ \left[ -k_1^3  {\cal F}\left[0,p_1\right]  -k_1^3 p_1 {\cal F} \left[1,p_1\right] +  \left(k_1 \left(k_2+k_3+k_4\right)- k_3 k_4-k_2 \left(k_3+k_4\right)\right) k_1 {\cal F}[2,p_1] + \right.  \right.  \nonumber \\ && \left. \left. \quad + \left(-k_2 k_3 k_4+k_1 \left(k_3 k_4+k_2 \left(k_3+k_4\right)\right) \right) {\cal F}[3,p_1] +k_2k_3k_4{\cal F}[4,p_1]  \right]   + \text{3 perm.} +   \right. \nonumber \\ && \left. \quad +k_1^3  {\cal F}\left[0,q_t\right]  +k_1^3 q_t {\cal F}\left[1,q_t\right] +  \left(k_3 k_4+k_2 \left(k_3+k_4\right)+k_1 \left(k_2+k_3+k_4\right)\right)   k_1 {\cal F}[2,q_t] + \right. \nonumber \\ && \left. \quad +  \left(k_2 k_3 k_4+k_1 \left(k_3 k_4+k_2 \left(k_3+k_4\right)\right)\right){\cal F}[3,q_t] +k_2k_3k_4{\cal F}[4,q_t]   \right] + \text{5 perm.} \nonumber
\end{eqnarray}
For shapes far from the flattened limit and for $\alpha \gg 1$ it becomes
\begin{eqnarray} \label{I3 nonflat}
&& \mathcal{I}_3 ^\text{non-flat}(k_1,k_2,k_3,k_4)  =  |c_s| \frac{\left( \q_1 \cdot \q_2 \right) \left(\q_3 \cdot \q_4 \right)  }{k_1^4}  \times  \\ && \quad \left[ \left[\frac{1 }{p_1^5} \left(k_1^3 (p_1-1) p_1^3+6 k_2 k_3 k_4 (p_1-4) +   \left(k_2+k_3+k_4\right) (p_1-2) p_1^2 k_1^2+\right. \right. \right. \nonumber \\ && \left.  \left. \left. \quad  +2 \left(k_3 k_4+k_2 \left(k_3+k_4\right)\right) k_1 (p_1-3) p_1 \right) + \text{3 perm.}\right]   -  \right. \nonumber \\ && \left. \quad  - \frac{1}{q_t^5}\left(k_1 q_t \left(k_1 q_t \left(k_1 q_t \left(q_t+1\right)+k_4 \left(q_t+2\right)\right)+k_3 \left(k_1 q_t \left(q_t+2\right)+2 k_4 \left(q_t+3\right)\right)\right) + \right. \right. \nonumber \\ && \left. \left. \quad + k_2 \left(k_1 q_t \left(k_1 q_t \left(q_t+2\right)+2 k_4 \left(q_t+3\right)\right)+2 k_3 \left(k_1 q_t \left(q_t+3\right)+3 k_4 \left(q_t+4\right)\right)\right)\right) \right]+ \text{5 perm.} \nonumber
\end{eqnarray}
\subsection{Exchanged diagram: equilateral limits}
In the equilateral limit, $k_1=k_2=k_3=k_4$, the different contributions to the trispectrum in the large-$\alpha$ limit are given by 
\begin{eqnarray}
{\cal T}_i^\text{equi}=  -   \left(\frac{1}{|c_{s}^{2}|}+1\right)^2  \frac{1}{k^3 2^9 |c_s|^2} {\cal K}_i^\text{equi}  \, ,
\end{eqnarray}
where
\begin{eqnarray} 
&& \label{K1 equi} {\cal K}_1^\text{equi}(k,k,k,k)=  -4 \times 144 \left(\frac{A}{c_s^2}\right)^2 b_{12}  k^3 |c_s|^6 \times \\ && \, \,  \left(\frac{4}{b_{12}^6}+\frac{4}{(b_{12}+2)^6}-\frac{8}{b_{12}^3 (b_{12}+2)^3}-b_{12} \frac{ 93 b_{12}^4-1048b_{12}^2+3600 }{64 \left(4-b_{12}^2\right)^3}\right)  +  (b_{12} \leftrightarrow \{b_{14},b_{13} \} )  \,,  \nonumber 
\end{eqnarray}
\begin{eqnarray}
&& \label{K2 equi} {\cal K}_2^\text{equi}(k,k,k,k)=\frac 3 2  \left(\frac{A}{|c_s^2|}\right) \frac{  k^3 |c_s|^4 }{ \left(b_{12}-2\right){}^3 b_{12}^5 \left(b_{12}+1\right){}^4 \left(b_{12}+2\right){}^6}  \\ &&  \, \, \left(262144+1441792 b_{12}+2867200 b_{12}^2+1548288 b_{12}^3-2899968 b_{12}^4-4988928 b_{12}^5-1304064 b_{12}^6+ \right. \nonumber \\ && \left. \, \,2437632 b_{12}^7+1527552 b_{12}^8-494976 b_{12}^9-407520 b_{12}^{10}+573312 b_{12}^{11}+920192 b_{12}^{12}+543296 b_{12}^{13}+\right. \nonumber \\ && \left. \, \, 43904 b_{12}^{14}-130766 b_{12}^{15}-67973 b_{12}^{16}-4424 b_{12}^{17} +211 b_{12}^{20}+2110 b_{12}^{19}+6394 b_{12}^{18} \right)  +  b_{12} \leftrightarrow \{b_{14},b_{13} \}  \, , \nonumber 
\end{eqnarray}
\begin{eqnarray}
&& \label{K3 equi} {\cal K}_3^\text{equi}(k,k,k,k)=-\frac 3 2 \left(\frac{A}{|c_s^2|}\right) \frac{k^3 |c_s|^4 }{ \left(b_{12}-2\right){}^3 b_{12}^5 \left(b_{12}+1\right){}^5 \left(b_{12}+2\right){}^6} \\ && \, \, \left(-262144-1441792 b_{12}-2375680 b_{12}^2+565248 b_{12}^3+5849088 b_{12}^4+5480448 b_{12}^5-443904 b_{12}^6-\right. \nonumber \\ && \left. \, \, 3396096 b_{12}^7-1380096 b_{12}^8+782976 b_{12}^9+973344 b_{12}^{10}+585216 b_{12}^{11}+409408 b_{12}^{12}+218368 b_{12}^{13}+ \right. \nonumber \\ && \left. \, \, 77 b_{12}^{20}+770 b_{12}^{19}+2342 b_{12}^{18}-1528 b_{12}^{17}-24475 b_{12}^{16}-47218 b_{12}^{15}+19168 b_{12}^{14} \right) +  b_{12} \leftrightarrow \{b_{14},b_{13} \} 
 \nonumber \, , 
 \end{eqnarray}
 \begin{eqnarray}
&&{\cal K}_4^\text{equi} (k,k,k,k)=  \frac{16 |c_s|^2  k^3 }{ \left(b_{12}-2\right){}^3 b_{12}^5 \left(b_{12}+1\right){}^5 \left(b_{12}+2\right){}^6}  \\ && \, \, \left(-524288-786432 b_{12}+1048576 b_{12}^2+2490368 b_{12}^3+196608 b_{12}^4-2555904 b_{12}^5-1449984 b_{12}^6+\right. \nonumber \\ && \left.  \, \, 1046016 b_{12}^7+732416 b_{12}^8-767104 b_{12}^9-516800 b_{12}^{10}+469152 b_{12}^{11}+478064 b_{12}^{12}+51656 b_{12}^{13}- \right. \nonumber \\ && \left. \, \, 435 b_{12}^{18}+3214 b_{12}^{17}+2984 b_{12}^{16}-32840 b_{12}^{15}+89780 b_{12}^{14} \right) +  b_{12} \leftrightarrow \{b_{14},b_{13} \} \, .
\nonumber \label{K4 equi}
\end{eqnarray}
For completeness I also give the integral form of $\cK_4$ after inserting the mode functions in eq. \eqref{K4} 
\begin{eqnarray} \label{K4 app}
&& \cK_4 (k_1,k_2,k_3,k_4)= - 16 |c_s|^2 \int_{\alpha}^{0} dz_1 \int_{\alpha}^{z_1} d z_2 \left[ (\q_1 \cdot \q_2 ) \left[  \frac{k_{12} }{k_4^5}(\q_3 \cdot \q_4 )    e^{p_2  z_1+p_1 z_2}   \times \right. \right. \nonumber \\ && \left. \left.  \quad \quad \left(k_4 \left(z_1 \left(-e^{\frac{2 k_3 z_1}{k_4}}+e^{2 z_1}+2\right)-e^{\frac{2 k_3 z_1}{k_4}}-e^{2 z_1}+2\right)+ \right. \right. \right. \nonumber \\ && \left. \left. \left. \quad \quad k_3 z_1 \left(z_1 \left(e^{\frac{2 k_3 z_1}{k_4}}+e^{2 z_1}+2\right)+e^{\frac{2 k_3 z_1}{k_4}}-e^{2 z_1}+2\right)\right) \times \right. \right. \nonumber \\ && \left. \left.  \quad \quad  \left( k_4 z_2 \left(k_1 \left(e^{\frac{2 k_{12} z_1}{k_4}}+e^{\frac{2 k_1 z_2}{k_4}}-e^{\frac{2 k_2 z_2}{k_4}}-e^{\frac{2 k_{12} z_2}{k_4}}+2\right)+ \right.  \right. \right. \right.  \nonumber \\ && \left. \left.  \left.   \left. \quad \quad  k_2 \left(e^{\frac{2 k_{12} z_1}{k_4}}-e^{\frac{2 k_1 z_2}{k_4}}+e^{\frac{2 k_2 z_2}{k_4}}-e^{\frac{2 k_{12} z_2}{k_4}}+2\right)\right)+ \right. \right. \right. \nonumber \\ && \left. \left.  \left.  \quad \quad  + k_1 k_2 z_2^2 \left(e^{\frac{2 k_{12} z_1}{k_4}}+e^{\frac{2 k_1 z_2}{k_4}}+e^{\frac{2 k_2 z_2}{k_4}}-e^{\frac{2 k_{12} z_2}{k_4}}+2\right)  +\right.  \right. \right. \nonumber \\ && \left. \left.  \left.  \quad \quad k_4^2 \left(e^{\frac{2 k_{12} z_1}{k_4}}-e^{\frac{2 k_1 z_2}{k_4}}-e^{\frac{2 k_2 z_2}{k_4}}-e^{\frac{2 k_{12} z_2}{k_4}}+2\right) \right) +\right.  \right. \nonumber \\ && \left. \left. \quad \quad  \text{2 perm.} \right] + \text{2 perm. }\right]  + \text{5 perm.} 
\end{eqnarray}

\bibliography{4pfrefs}

\providecommand{\href}[2]{#2}\begingroup\raggedright\begin{thebibliography}{10}

\bibitem{Freese:1990rb}
K.~Freese, J.~A. Frieman and A.~V. Olinto, \emph{{Natural inflation with pseudo
  - Nambu-Goldstone bosons}},
  \href{http://dx.doi.org/10.1103/PhysRevLett.65.3233}{\emph{Phys. Rev. Lett.}
  {\bf 65} (1990) 3233--3236}.

\bibitem{Silverstein:2008sg}
E.~Silverstein and A.~Westphal, \emph{{Monodromy in the CMB: Gravity Waves and
  String Inflation}},
  \href{http://dx.doi.org/10.1103/PhysRevD.78.106003}{\emph{Phys. Rev.} {\bf
  D78} (2008) 106003}, [\href{https://arxiv.org/abs/0803.3085}{{\tt
  0803.3085}}].

\bibitem{Kaloper:2008fb}
N.~Kaloper and L.~Sorbo, \emph{{A Natural Framework for Chaotic Inflation}},
  \href{http://dx.doi.org/10.1103/PhysRevLett.102.121301}{\emph{Phys. Rev.
  Lett.} {\bf 102} (2009) 121301}, [\href{https://arxiv.org/abs/0811.1989}{{\tt
  0811.1989}}].

\bibitem{Akrami:2018odb}
{\scshape Planck} collaboration, Y.~Akrami et~al., \emph{{Planck 2018 results.
  X. Constraints on inflation}},  \href{https://arxiv.org/abs/1807.06211}{{\tt
  1807.06211}}.

\bibitem{Hebecker:2019vyf}
A.~Hebecker and P.~Henkenjohann, \emph{{Gauge and gravitational instantons:
  From 3-forms and fermions to Weak Gravity and flat axion potentials}},
  \href{http://dx.doi.org/10.1007/JHEP09(2019)038}{\emph{JHEP} {\bf 09} (2019)
  038}, [\href{https://arxiv.org/abs/1906.07728}{{\tt 1906.07728}}].

\bibitem{Cremonini:2010ua}
S.~Cremonini, Z.~Lalak and K.~Turzynski, \emph{{Strongly Coupled Perturbations
  in Two-Field Inflationary Models}},
  \href{http://dx.doi.org/10.1088/1475-7516/2011/03/016}{\emph{JCAP} {\bf 1103}
  (2011) 016}, [\href{https://arxiv.org/abs/1010.3021}{{\tt 1010.3021}}].

\bibitem{Achucarro:2010jv}
A.~Achucarro, J.-O. Gong, S.~Hardeman, G.~A. Palma and S.~P. Patil, \emph{{Mass
  hierarchies and non-decoupling in multi-scalar field dynamics}},
  \href{http://dx.doi.org/10.1103/PhysRevD.84.043502}{\emph{Phys. Rev.} {\bf
  D84} (2011) 043502}, [\href{https://arxiv.org/abs/1005.3848}{{\tt
  1005.3848}}].

\bibitem{Achucarro:2012yr}
A.~Achucarro, V.~Atal, S.~Cespedes, J.-O. Gong, G.~A. Palma and S.~P. Patil,
  \emph{{Heavy fields, reduced speeds of sound and decoupling during
  inflation}}, \href{http://dx.doi.org/10.1103/PhysRevD.86.121301}{\emph{Phys.
  Rev.} {\bf D86} (2012) 121301}, [\href{https://arxiv.org/abs/1205.0710}{{\tt
  1205.0710}}].

\bibitem{Cespedes:2012hu}
S.~Cespedes, V.~Atal and G.~A. Palma, \emph{{On the importance of heavy fields
  during inflation}},
  \href{http://dx.doi.org/10.1088/1475-7516/2012/05/008}{\emph{JCAP} {\bf 1205}
  (2012) 008}, [\href{https://arxiv.org/abs/1201.4848}{{\tt 1201.4848}}].

\bibitem{Renaux-Petel:2015mga}
S.~Renaux-Petel and K.~Turzyński, \emph{{Geometrical Destabilization of
  Inflation}},
  \href{http://dx.doi.org/10.1103/PhysRevLett.117.141301}{\emph{Phys. Rev.
  Lett.} {\bf 117} (2016) 141301},
  [\href{https://arxiv.org/abs/1510.01281}{{\tt 1510.01281}}].

\bibitem{Brown:2017osf}
A.~R. Brown, \emph{{Hyperbolic Inflation}},
  \href{http://dx.doi.org/10.1103/PhysRevLett.121.251601}{\emph{Phys. Rev.
  Lett.} {\bf 121} (2018) 251601},
  [\href{https://arxiv.org/abs/1705.03023}{{\tt 1705.03023}}].

\bibitem{Mizuno:2017idt}
S.~Mizuno and S.~Mukohyama, \emph{{Primordial perturbations from inflation with
  a hyperbolic field-space}},
  \href{http://dx.doi.org/10.1103/PhysRevD.96.103533}{\emph{Phys. Rev.} {\bf
  D96} (2017) 103533}, [\href{https://arxiv.org/abs/1707.05125}{{\tt
  1707.05125}}].

\bibitem{Christodoulidis:2018qdw}
P.~Christodoulidis, D.~Roest and E.~I. Sfakianakis, \emph{{Angular inflation in
  multi-field $\alpha$-attractors}},
  \href{http://dx.doi.org/10.1088/1475-7516/2019/11/002}{\emph{JCAP} {\bf 1911}
  (2019) 002}, [\href{https://arxiv.org/abs/1803.09841}{{\tt 1803.09841}}].

\bibitem{Garcia-Saenz:2018ifx}
S.~Garcia-Saenz, S.~Renaux-Petel and J.~Ronayne, \emph{{Primordial fluctuations
  and non-Gaussianities in sidetracked inflation}},
  \href{http://dx.doi.org/10.1088/1475-7516/2018/07/057}{\emph{JCAP} {\bf 1807}
  (2018) 057}, [\href{https://arxiv.org/abs/1804.11279}{{\tt 1804.11279}}].

\bibitem{Bravo:2019xdo}
R.~Bravo, G.~A. Palma and S.~Riquelme, \emph{{A Tip for Landscape Riders:
  Multi-Field Inflation Can Fulfill the Swampland Distance Conjecture}},
  \href{http://dx.doi.org/10.1088/1475-7516/2020/02/004}{\emph{JCAP} {\bf 2002}
  (2020) 004}, [\href{https://arxiv.org/abs/1906.05772}{{\tt 1906.05772}}].

\bibitem{Aragam:2019khr}
V.~Aragam, S.~Paban and R.~Rosati, \emph{{Multi-field Inflation in High-Slope
  Potentials}},  \href{https://arxiv.org/abs/1905.07495}{{\tt 1905.07495}}.

\bibitem{Garcia-Saenz:2019njm}
S.~Garcia-Saenz, L.~Pinol and S.~Renaux-Petel, \emph{{Revisiting
  non-Gaussianity in multifield inflation with curved field space}},
  \href{http://dx.doi.org/10.1007/JHEP01(2020)073}{\emph{JHEP} {\bf 01} (2020)
  073}, [\href{https://arxiv.org/abs/1907.10403}{{\tt 1907.10403}}].

\bibitem{Bjorkmo:2019aev}
T.~Bjorkmo and M.~C.~D. Marsh, \emph{{Hyperinflation generalised: from its
  attractor mechanism to its tension with the ‘swampland conditions’}},
  \href{http://dx.doi.org/10.1007/JHEP04(2019)172}{\emph{JHEP} {\bf 04} (2019)
  172}, [\href{https://arxiv.org/abs/1901.08603}{{\tt 1901.08603}}].

\bibitem{Bjorkmo:2019fls}
T.~Bjorkmo, \emph{{Rapid-Turn Inflationary Attractors}},
  \href{http://dx.doi.org/10.1103/PhysRevLett.122.251301}{\emph{Phys. Rev.
  Lett.} {\bf 122} (2019) 251301},
  [\href{https://arxiv.org/abs/1902.10529}{{\tt 1902.10529}}].

\bibitem{Chakraborty:2019dfh}
D.~Chakraborty, R.~Chiovoloni, O.~Loaiza-Brito, G.~Niz and I.~Zavala,
  \emph{{Fat Inflatons, Large Turns and the $\eta$-problem}},
  \href{http://dx.doi.org/10.1088/1475-7516/2020/01/020}{\emph{JCAP} {\bf 2001}
  (2020) 020}, [\href{https://arxiv.org/abs/1908.09797}{{\tt 1908.09797}}].

\bibitem{Christodoulidis:2019jsx}
P.~Christodoulidis, D.~Roest and E.~I. Sfakianakis, \emph{{Scaling attractors
  in multi-field inflation}},
  \href{http://dx.doi.org/10.1088/1475-7516/2019/12/059}{\emph{JCAP} {\bf 1912}
  (2019) 059}, [\href{https://arxiv.org/abs/1903.06116}{{\tt 1903.06116}}].

\bibitem{Langlois:2008mn}
D.~Langlois and S.~Renaux-Petel, \emph{{Perturbations in generalized
  multi-field inflation}},
  \href{http://dx.doi.org/10.1088/1475-7516/2008/04/017}{\emph{JCAP} {\bf 0804}
  (2008) 017}, [\href{https://arxiv.org/abs/0801.1085}{{\tt 0801.1085}}].

\bibitem{Cheung:2007st}
C.~Cheung, P.~Creminelli, A.~L. Fitzpatrick, J.~Kaplan and L.~Senatore,
  \emph{{The Effective Field Theory of Inflation}},
  \href{http://dx.doi.org/10.1088/1126-6708/2008/03/014}{\emph{JHEP} {\bf 03}
  (2008) 014}, [\href{https://arxiv.org/abs/0709.0293}{{\tt 0709.0293}}].

\bibitem{Achucarro:2010da}
A.~Achucarro, J.-O. Gong, S.~Hardeman, G.~A. Palma and S.~P. Patil,
  \emph{{Features of heavy physics in the CMB power spectrum}},
  \href{http://dx.doi.org/10.1088/1475-7516/2011/01/030}{\emph{JCAP} {\bf 1101}
  (2011) 030}, [\href{https://arxiv.org/abs/1010.3693}{{\tt 1010.3693}}].

\bibitem{Achucarro:2012sm}
A.~Achucarro, J.-O. Gong, S.~Hardeman, G.~A. Palma and S.~P. Patil,
  \emph{{Effective theories of single field inflation when heavy fields
  matter}}, \href{http://dx.doi.org/10.1007/JHEP05(2012)066}{\emph{JHEP} {\bf
  05} (2012) 066}, [\href{https://arxiv.org/abs/1201.6342}{{\tt 1201.6342}}].

\bibitem{Anber2010}
M.~M. Anber and L.~Sorbo, \emph{Naturally inflating on steep potentials through
  electromagnetic dissipation},
  \href{http://dx.doi.org/10.1103/PhysRevD.81.043534}{\emph{Phys. Rev.} {\bf
  D81} (2010) 043534}, [\href{https://arxiv.org/abs/0908.4089}{{\tt
  0908.4089}}].

\bibitem{Barnaby2011}
N.~Barnaby, R.~Namba and M.~Peloso, \emph{Phenomenology of a pseudo-scalar
  inflaton: Naturally large nongaussianity},
  \href{http://dx.doi.org/10.1088/1475-7516/2011/04/009}{\emph{JCAP} {\bf 1104}
  (2011) 009}, [\href{https://arxiv.org/abs/1102.4333}{{\tt 1102.4333}}].

\bibitem{Ferreira2014}
R.~Z. Ferreira and M.~S. Sloth, \emph{Universal constraints on axions from
  inflation}, \href{http://dx.doi.org/10.1007/JHEP12(2014)139}{\emph{JHEP} {\bf
  12} (2014) 139}, [\href{https://arxiv.org/abs/1409.5799}{{\tt 1409.5799}}].

\bibitem{Ferreira2017}
R.~Z. Ferreira and A.~Notari, \emph{Thermalized axion inflation},
  \href{http://dx.doi.org/10.1088/1475-7516/2017/09/007}{\emph{JCAP} {\bf 1709}
  (2017) 007}, [\href{https://arxiv.org/abs/1706.00373}{{\tt 1706.00373}}].

\bibitem{Garcia-Saenz:2018vqf}
S.~Garcia-Saenz and S.~Renaux-Petel, \emph{{Flattened non-Gaussianities from
  the effective field theory of inflation with imaginary speed of sound}},
  \href{http://dx.doi.org/10.1088/1475-7516/2018/11/005}{\emph{JCAP} {\bf 1811}
  (2018) 005}, [\href{https://arxiv.org/abs/1805.12563}{{\tt 1805.12563}}].

\bibitem{Bjorkmo:2019qno}
T.~Bjorkmo, R.~Z. Ferreira and M.~C.~D. Marsh, \emph{{Mild Non-Gaussianities
  under Perturbative Control from Rapid-Turn Inflation Models}},
  \href{http://dx.doi.org/10.1088/1475-7516/2019/12/036}{\emph{JCAP} {\bf 1912}
  (2019) 036}, [\href{https://arxiv.org/abs/1908.11316}{{\tt 1908.11316}}].

\bibitem{Fumagalli:2019noh}
J.~Fumagalli, S.~Garcia-Saenz, L.~Pinol, S.~Renaux-Petel and J.~Ronayne,
  \emph{{Hyper-Non-Gaussianities in Inflation with Strongly Nongeodesic
  Motion}}, \href{http://dx.doi.org/10.1103/PhysRevLett.123.201302}{\emph{Phys.
  Rev. Lett.} {\bf 123} (2019) 201302},
  [\href{https://arxiv.org/abs/1902.03221}{{\tt 1902.03221}}].

\bibitem{Chen:2006nt}
X.~Chen, M.-x. Huang, S.~Kachru and G.~Shiu, \emph{{Observational signatures
  and non-Gaussianities of general single field inflation}},
  \href{http://dx.doi.org/10.1088/1475-7516/2007/01/002}{\emph{JCAP} {\bf 0701}
  (2007) 002}, [\href{https://arxiv.org/abs/hep-th/0605045}{{\tt
  hep-th/0605045}}].

\bibitem{Chen:2009bc}
X.~Chen, B.~Hu, M.-x. Huang, G.~Shiu and Y.~Wang, \emph{{Large Primordial
  Trispectra in General Single Field Inflation}},
  \href{http://dx.doi.org/10.1088/1475-7516/2009/08/008}{\emph{JCAP} {\bf 0908}
  (2009) 008}, [\href{https://arxiv.org/abs/0905.3494}{{\tt 0905.3494}}].

\bibitem{Calzetta1987}
E.~Calzetta and B.~L. Hu, \emph{Closed-time-path functional formalism in curved
  spacetime: Application to cosmological back-reaction problems},
  \href{http://dx.doi.org/10.1103/physrevd.35.495}{\emph{Physical Review D}
  {\bf 35} (jan, 1987) 495--509}.

\bibitem{Weinberg2005}
S.~Weinberg, \emph{Quantum contributions to cosmological correlations},
  \href{https://arxiv.org/abs/hep-th/0506236v1}{{\tt hep-th/0506236v1}}.

\bibitem{Ferreira:2015omg}
R.~Z. Ferreira, J.~Ganc, J.~Noreña and M.~S. Sloth, \emph{{On the validity of
  the perturbative description of axions during inflation}},
  \href{http://dx.doi.org/10.1088/1475-7516/2016/10/E01,
  10.1088/1475-7516/2016/04/039}{\emph{JCAP} {\bf 1604} (2016) 039},
  [\href{https://arxiv.org/abs/1512.06116}{{\tt 1512.06116}}].

\bibitem{Germani:2018jgr}
C.~Germani and I.~Musco, \emph{{Abundance of Primordial Black Holes Depends on
  the Shape of the Inflationary Power Spectrum}},
  \href{http://dx.doi.org/10.1103/PhysRevLett.122.141302}{\emph{Phys. Rev.
  Lett.} {\bf 122} (2019) 141302},
  [\href{https://arxiv.org/abs/1805.04087}{{\tt 1805.04087}}].

\bibitem{Meerburg:2019qqi}
P.~D. Meerburg et~al., \emph{{Primordial Non-Gaussianity}},
  \href{https://arxiv.org/abs/1903.04409}{{\tt 1903.04409}}.

\bibitem{Akrami:2019izv}
{\scshape Planck} collaboration, Y.~Akrami et~al., \emph{{Planck 2018 results.
  IX. Constraints on primordial non-Gaussianity}},
  \href{https://arxiv.org/abs/1905.05697}{{\tt 1905.05697}}.

\bibitem{Senatore:2009gt}
L.~Senatore, K.~M. Smith and M.~Zaldarriaga, \emph{{Non-Gaussianities in Single
  Field Inflation and their Optimal Limits from the WMAP 5-year Data}},
  \href{http://dx.doi.org/10.1088/1475-7516/2010/01/028}{\emph{JCAP} {\bf 1001}
  (2010) 028}, [\href{https://arxiv.org/abs/0905.3746}{{\tt 0905.3746}}].

\bibitem{Meerburg:2009fi}
P.~D. Meerburg, J.~P. van~der Schaar and M.~G. Jackson, \emph{{Bispectrum
  signatures of a modified vacuum in single field inflation with a small speed
  of sound}},
  \href{http://dx.doi.org/10.1088/1475-7516/2010/02/001}{\emph{JCAP} {\bf 1002}
  (2010) 001}, [\href{https://arxiv.org/abs/0910.4986}{{\tt 0910.4986}}].

\bibitem{Seery:2005wm}
D.~Seery and J.~E. Lidsey, \emph{{Primordial non-Gaussianities in single field
  inflation}},
  \href{http://dx.doi.org/10.1088/1475-7516/2005/06/003}{\emph{JCAP} {\bf 0506}
  (2005) 003}, [\href{https://arxiv.org/abs/astro-ph/0503692}{{\tt
  astro-ph/0503692}}].

\end{thebibliography}\endgroup
\bibliographystyle{JHEP}
\end{document}